\documentclass[twocolumn,trackchanges]{aastex7}
\usepackage{upgreek}
\usepackage[T1]{fontenc}

\shorttitle{Exorings with JWST}

\begin{document}

\title{On the Detection of Exorings in Reflected Light with JWST NIRCam}

\correspondingauthor{Rachel Bowens-Rubin}
\email{rbowru@umich.edu}

\author[0000-0001-5831-9530]
{Rachel Bowens-Rubin}
\email{}
\affiliation{Department of Astronomy, University of Michigan, Ann Arbor, MI 48109, USA}
\affiliation{Eureka Scientific Inc., 2542 Delmar Ave., Suite 100, Oakland, CA 94602, USA}

\author[0000-0002-9521-9798]{Mary Anne Limbach}
\email{}
\affiliation{Department of Astronomy, University of Michigan, Ann Arbor, MI 48109, USA}

\author[0009-0004-6318-6874]{Sam Hopper}
\email{ehop@umich.edu}
\affiliation{Department of Astronomy, University of Michigan, Ann Arbor, MI 48109, USA}

\author[0009-0008-2252-7969]{Klaus Subbotina Stephenson}
\email{klausss@uvic.ca}
\affiliation{Department of Physics and Astronomy University of Victoria 3800 Finnerty Road Elliot Building, Room 207. Victoria BC V8P 5C2}
\affiliation{Department of Astronomy \& Astrophysics, University of California, Santa Cruz, 1156 High St, Santa Cruz, CA 95064, USA}

\author[0009-0004-7835-3356]{Matson Garza}
\email{matgarza@mit.edu}
\affiliation{Department of Physics, Massachusetts Institute of Technology, 77 Massachusetts Avenue, Cambridge, MA 02139, USA}

\author[0000-0001-5834-9588]{Leigh N. Fletcher}
\email{leigh.fletcher@leicester.ac.uk}
\affiliation{School of Physics and Astronomy, University of Leicester, University Rd, Leicester LE1 7RH, United Kingdom}

\author[0000-0002-8592-0812]{Matthew Hedman}
\email{mhedman@uidaho.edu}
\affiliation{Department of Physics, 875 Perimeter Drive MS 0903, Moscow, ID 83844}

\begin{abstract}

When directly imaging a cold giant exoplanet hosting a ring system, 
the reflected light from the rings can outshine the planet’s thermal emission and reflected-light in the near-infrared.
Consequently, an exoring may be detectable at a significantly lower contrasts than is required to image the exoplanet itself.
Here we investigate the detectability of exorings in near-infrared reflected light using NIRCam coronagraphy \texttt{PanCAKE} simulations of two nearby mature stars, Proxima Centauri and Tau Ceti. 
Under the most favorable assumptions, we find JWST 2$\mu m$ NIRCam coronagraphy (F200W + MASK335R) is capable of detecting an exoring system with a radius of 2.8 times that of Saturn's A-ring for planets on an orbit with $a = 1.3-1.9\,$AU.  
Broader simulations indicate that NIRCam can probe large planetary ring systems around mature exoplanets comparable in size to circumplanetary disks, which can reach up to 1000 times the radius of Saturn’s A-ring.
These results suggest that NIRCam F200W coronagraphy could serendipitously detect large exorings in reflected light under the right conditions.
A combined analysis of F200W coronagraphic observations of confirmed exoplanets could provide the first empirical constraints on the occurrence rate of large exorings.
Confirming the existence and frequency of exorings spanning the scale between circumplanetary disks and the rings of the Solar System giant planet could offer new insight into the formation, evolution, and architecture of planetary systems.

\end{abstract}


\section{Introduction} \label{sec:intro}

Ringed exoplanets offer a unique opportunity to study planetary formation, dynamical evolution, and system architectures. Several theoretical and observational studies propose mechanisms by which rings may form around exoplanets, including collisions between minor bodies \citep{2015NatGe...8..686H}, tidal disruption or stripping of a minor body \citep{2010Natur.468..943C,2022Sci...377.1285W}, the incomplete dissipation of a primordial circumplanetary disk \citep{2005Icar..175..111B}, and material ejection from an active moon \citep{1979Sci...204..951S}. One of the major open questions about Saturn concerns the age of its rings and the rate to which they are replenished. 
Understanding the census of planetary rings beyond the Solar System could provide valuable constraints on these processes for a diversity of worlds, improve our understanding of planetary system demographics, and place the frequency of ringed exoplanets in the broader context.

The presence of a ring system can significantly affect the observable properties of an exoplanet, influencing their inferred radii, densities, and albedos \citep{ArnoldandSchneider2006, BarnesandFortney2004}. 
For example, the anomalously large radii of the so-called “super-puff” planets HIP\,41378\,f and Kepler\,90\,g relative to their masses have been hypothesized to result from the presence of extended ring systems \citep{2021AJ....161..202L,2025ApJ...980...39L, Akinsanmi2020}. A number of other exoplanets with well-characterized mass and radius measurements also fall into the super-puff category \citep{2020AJ....159..131P}. 
If rings are indeed responsible for the observed low densities of these planets, this would imply that ringed exoplanets may be somewhat common, particularly given the significant number of long-period transiting planets with measured masses that appear underdense. 

In addition to influencing bulk properties, rings can also affect a planet’s atmospheric composition and thermal structure. The material from the rings may fall onto the planet, delivering exogenic species such as oxygen-bearing compounds to the upper atmosphere. This influx can initiate photochemical reactions involving oxygen and carbon, as observed in Saturn’s atmosphere from Cassini data \citep{Waite2018}. Ring particles may also follow magnetic field lines and deposit in specific regions, altering the ionospheric composition at mid-latitudes through localized "ring rain" \citep{ODonoghue2016}. Furthermore, rings can cast shadows on the planetary atmosphere, creating temperature gradients and localized cooling in the stratosphere. These shadowing effects have been shown to influence atmospheric dynamics on Saturn, leading to banded temperature structures and circulation patterns that redistribute heat \citep{Fletcher2018}. Together, these processes demonstrate that rings may alter not only how exoplanets appear in observations but also their atmospheric chemistry and climate.

\begin{figure*}
\gridline{\fig{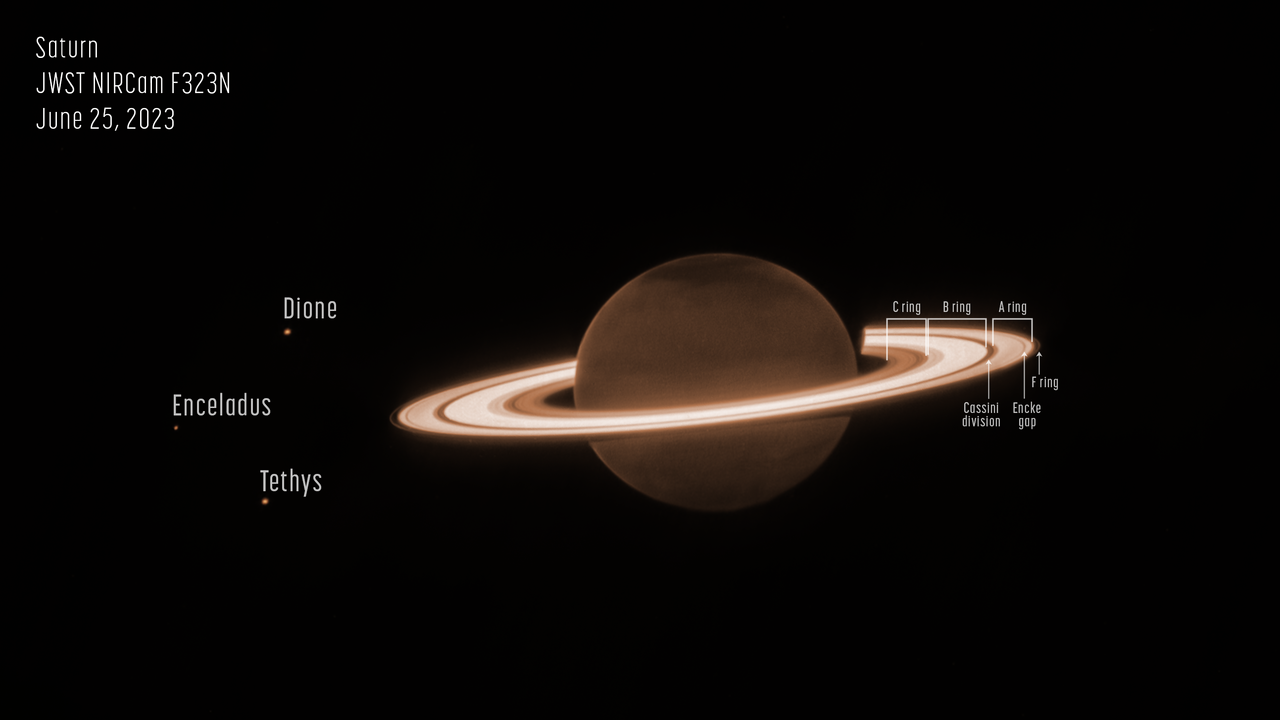}{0.6\textwidth}{(a) Saturn NIRCam F323N}
\fig{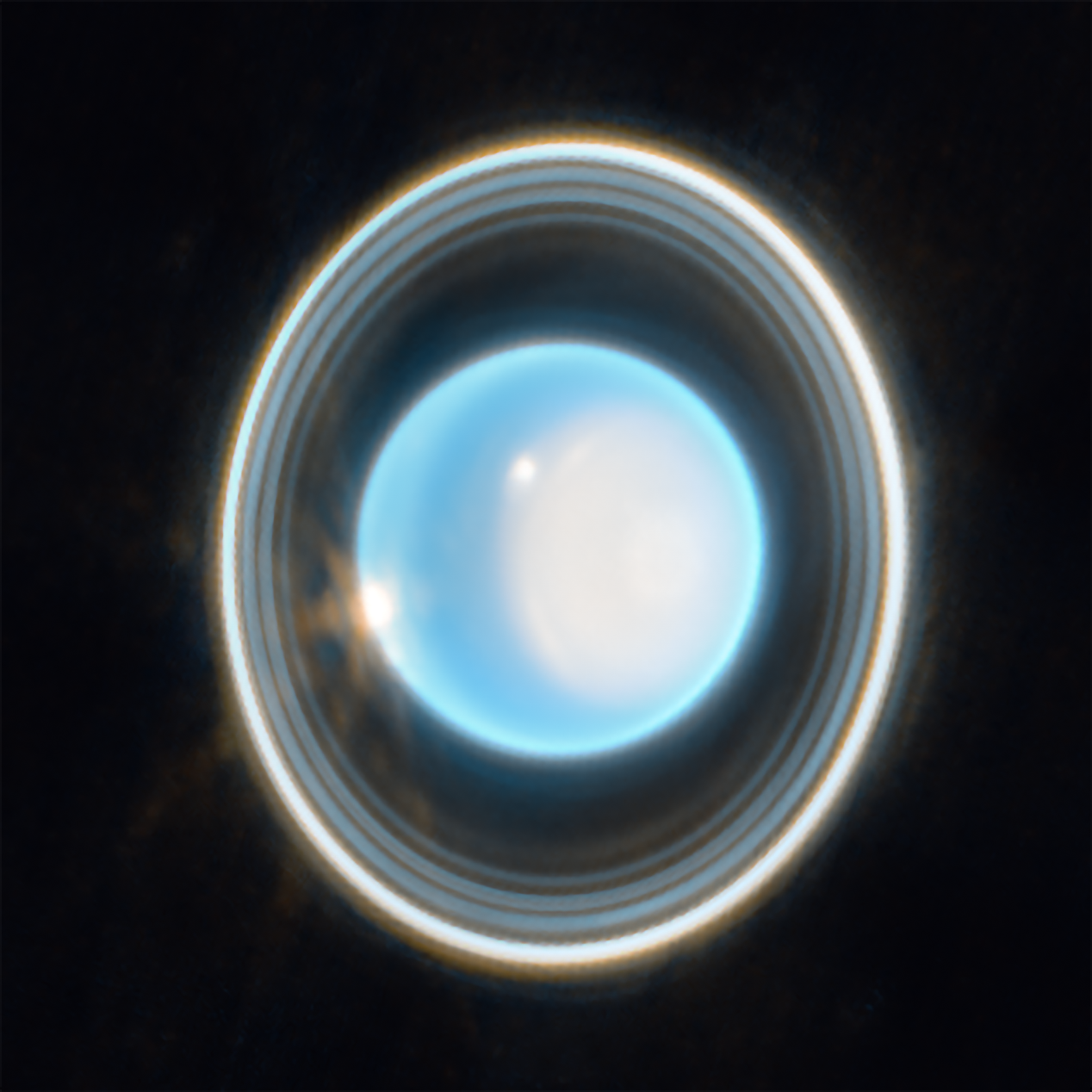}{0.34\textwidth}{(b)Uranus NIRCam F140M (Cyan) \& F300M (Orange) Composite }}
\caption{\textbf{The rings of Saturn and Uranus as imaged by JWST NIRCam}. These images provide an empirical example in which the flux of the ring system is comparable to, or even exceeds, that of the planet when observed with NIRCam.  The image of Saturn was created using data from JWST GTO 1247.\footnote{GTO 1247: \url{https://www.stsci.edu/jwst/science-execution/program-information?id=1247}; Image Credit: NASA, ESA, CSA, Matthew Tiscareno (SETI Institute), Matthew Hedman (University of Idaho), Maryame El Moutamid (Cornell University), Mark Showalter (SETI Institute), Leigh Fletcher (University of Leicester), Heidi Hammel (AURA)} The image of Uranus was created using data from JWST DD 2739.\footnote{DD 2739: \url{https://www.stsci.edu/jwst-program-info/program/?program=2739}; Image Credit: Joseph DePasquale (STScI); NASA, ESA, CSA, STScI}}
\label{fig:planetswithJWST}
\end{figure*}

While an exoring system around a mature exoplanet has yet been conclusively confirmed, 
circumplanetary disks around young accreting exoplanets have been observed. Direct imaging detections of circumplanetary disks have been reported in several systems, including J1407b \citep{2015ApJ...800..126K}, PDS~70c \citep{2021ApJ...916L...2B}, AS~209 \citep{2022ApJ...934L..20B}, HD~100546b \citep{2013ApJ...766L...1Q}, and YSES-1b \citep{Hoch2025}. While circumplanetary disks trace the initial conditions of material accreting onto forming giant planets and the early stages of satellite formation \citep{2002AJ....124.3404C,2005Icar..175..111B,2018ApJ...868L..13S}, exorings detected around mature planets can offer similar insight into the dynamical evolution of planetary systems, including tidal interactions, satellite disruption, or migration-driven instabilities that reshape or destroy moon systems over time \citep{2010Natur.468..943C,2019AJ....157...30H}.

In reflected-light direct imaging, planetary rings can contribute significantly to the total brightness of a system. Because they can have a larger surface area and higher albedo than the planet, rings may outshine the planet itself under certain viewing geometries. In the Solar System, all four giant planets are known to host rings \citep{1977Natur.267..328E,1986Natur.319..636H,1989Sci...246.1422S}, with Saturn’s system being the most prominent. At near-infrared wavelengths, particularly where methane absorption bands reduce the planet’s flux, Saturn’s rings can appear nearly 100 times brighter than the planet. This effect is evident in recent JWST/NIRCam narrow-band imaging of Saturn with the F323N filter (see Figure~\ref{fig:planetswithJWST}). Similar behavior is observed in medium-band NIRCam filters, where the rings of both Saturn and Uranus can appear as bright or brighter than the planets themselves \citep{Hedman2025}.

Due to the high reflectivity of icy rings, the contrasts needed to detect exorings in reflected light can be orders of magnitude less than what would be required for reflected light or thermal emission imaging of mature (cold) giant planets. 
Moreover, targeting the light reflected off a planetary ring system may open the discovery space towards finding mature sub-Jupiter mass exoplanets that would otherwise be too cold/small/non-reflective to be directly detected by imaging. 

These characteristics raise a compelling question: could rings around cold/mature exoplanets in the nearest systems be detectable in reflected light with 
JWST NIRCam coronagraphy \citep{2018SPIE10698E..09P, Girard2022}?
In this work, we explore such a possible search by simulating the contrast of JWST NIRCam for two nearby systems and applying simplified ring models. 
In Section \ref{sec:methods}, we describe our methods for simulating the JWST contrast and understanding these contrast measurements with respect to ring-radius constraints. In Section \ref{sec:results}, we show the results of simulating the Proxima Centauri and Tau Ceti systems. Section~\ref{sec:discussion} presents our findings and explores future prospects for exoring detection, while Section~\ref{sec:conclusions} provides a summary of our results.

\section{Methods \label{sec:methods}}

\subsection{Simulating JWST NIRCam Coronagraphy Contrast}

We present simulations of two example cases of mature, nearby systems with different stellar types:  Proxima Centauri and Tau Ceti. Proxima Centauri is an M5.5V star  \citep{Bessell1991} of age 4.85 Gyr \citep{Kervella2003} at a distance of 1.3pc \citep{GaiaEDR3}.  Tau Ceti is a G8V star \citep{Keenan1989} with an age of 8 - 10 Gyr \citep{Tang2011} at 3.7 pc \citep{GaiaEDR3}.  Both of these systems have multiple confirmed exoplanets or candidates that have been identified through radial velocity \citep{Anglada-Escudé2016, Damasso2020, Dumusque2012, Faria2022, Feng2017}.

We simulated the NIRCam coronagraphy contrast performance using the Pandeia Coronagraphy Advanced Kit for Extractions (\texttt{PanCAKE}) python package\footnote{Pandeia Coronagraphy Advanced Kit for Extractions; \url{ https://github.com/spacetelescope/pandeia-coronagraphy}} (\cite{Girard2018}, \citealt{Perrin2018}, \citealt{CarterPancake}). In order to represent a case for a feasible future observation of a nearby brightstar, we replicated the observing configurations for previously performed NIRCam coronagraphy observations of nearby bright stars. 
For Proxima Centauri, we replicated the NIRCam configuration used to observe the nearby M-dwarf systems in the GO 6122 program in F200W/F444W \citep{CoolKidsprop}: SUB320/MEDIUM8 readout, 8 groups/int, 36 int/exp, and 3041s total time.
For Tau Ceti, we replicated the NIRCam configuration used to perform an observation of Fomalhaut in F356W in the GO 1193 program \citep{GO1193prop, Ygouf2024}: SUB320/RAPID readout, 3 groups/int, 105 ints/exp, and 451s total time. 

 We simulated the NIRCam coronagraphy observations of Proxima Centauri and Tau Ceti using all the combinations of masks and filters available in \texttt{PanCAKE}.  
Five masks were simulated: MASK210R, MASK335R, MASK430R, MASKLWB, and MASKSWB. The following filters were simulated in combination with their allowed masks: F164N, F182M, F187N, F200W, F210M, F212N, F250M, F277W, F300M, F322W2, F323N, F335M, F356W, F360M, F405N, F410M, F430M, F444W, F460M, F466N, F470N, F480M. 

The coronagraphs available with NIRCam cannot operate at shorter wavelengths ($<1.64\mu$m) because the optical wedges that project the occulting masks onto the detectors introduce chromatic aberrations, and the antireflective coating on the coronagraph substrate exhibits poor throughput in this regime. This limitation is unfortunate, as shorter wavelengths are likely optimal for detecting exorings in reflected light.

We modeled the PSF subtraction as angular differential imaging (ADI) \& reference differential imaging (RDI) with a 5-point-diamond dither on the reference star and a roll angle of 10 degrees. We assumed a perfect reference star match by inputting the target star properties as the reference star, following the steps of the  \texttt{PanCAKE} basic tutorial.\footnote{The \texttt{PanCAKE} tutorial referenced here can be accessed at  \url{https://aarynncarter.com/PanCAKE/notebooks/pancake_basic_tutorial.html}} 
Due to the internal limits of the \texttt{PanCAKE} software, the contrast limits were only simulated out to a separation of 1.46 arcsec. However, the field of view of the SUB320 array is 10 arcsec in the short wavelength filters,\footnote{Field of view of the SUB320 array with MASK335R is listed in Table 2 of the JDocs accessed at \url{https://jwst-docs.stsci.edu/jwst-near-infrared-camera/nircam-instrumentation/nircam-detector-overview/nircam-detector-subarrays}} so we estimated the contrasts to a 5-arcsec radius by extrapolating using a linear fit of the final 3 points of the \texttt{PanCAKE} simulation.  This extrapolation represents a pessimistic scenario, as if the best background limit was achieved by 1.46 arcsec. However, on-sky results have shown that improved background limits can be achieved beyond 1.5 arcsec of  separation (see NIRCam F444W+MASK335R contrast curves in \citealt{Carter2023}  and \citealt{Bowens-Rubin2025} examples which reach the best background limits beyond 2.5 arcsec).

We do not consider the use of NIRCam's direct imaging mode (no coronagraph) in this study, as the instrument generally saturates at the small angular separations corresponding to the projected locations of the optimal location for observing ring systems around nearby stars. This saturation renders observations of exorings infeasible at the spatial scales of interest. We focus this analysis on NIRCam's coronagraphic mode only.

\begin{figure*}
\begin{center}
\includegraphics[width=0.76\textwidth]{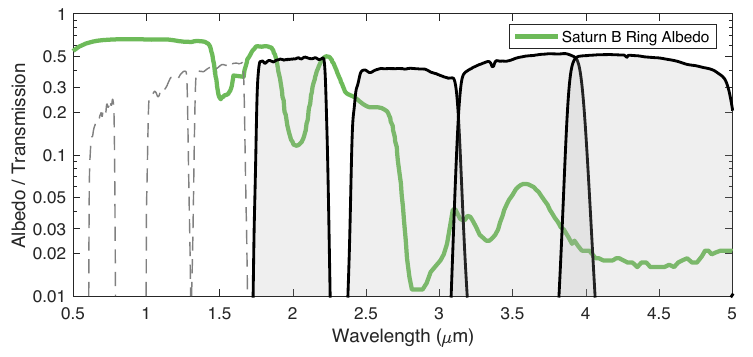}
\end{center}
\caption{\textbf{The albedo spectrum of Saturn’s B-ring across the visible and near-infrared.} The measurement of the albedo of Saturn's rings 
\textit{(green)} is plotted alongside the wavelength ranges of the JWST NIRCam wide-band filters. The filters unavailable for coronagraphy (F070W, F115W, F150W) are shown in dotted lines. The filters availble for corongraphy (F200W, F277W, F356W and F444W) are plotted with the thick black lines and gray shading.  Saturn's rings show a reasonably high reflectivity at $\lesssim2\mu m$ but drop significantly at longer wavelengths.}
\label{fig:AlbedoAndFilters}
\end{figure*}

\subsection{Converting Simulated Contrast to Ring Size Estimates \label{sec:methods-rings}}

The primary observational limitation in detecting reflected light from planetary rings is the contrast between the rings and the host star. The reflected flux scales proportionally with the incident stellar flux, thus the detectability of planetary rings is independent of the type of star the planet orbits. Ring systems have lower (more favorable) ring-to-star contrasts when they are closer to their star, larger in area, highly reflective, and in an optimal geometric configuration to reflect light at the observer.

To examine the set of possible exoplanet ring systems that could be detectable from our simulated contrast curves, we adopted the following assumptions:

\begin{itemize}
\item The ring is observed in a face-on  configuration, reflecting the maximum amount of light towards the observer.
\item The rings do not contribute flux from thermal emission; all detected flux arises from stellar light reflected by the ring. 
\item The optical depth of the ring is high throughout the full area of the ring . 
\item The thermal emission and reflected light from the planet contributes negligibly to the total flux compared to the reflected light from the ring.

\end{itemize}

We considered two cases for the ring's reflectivity:
\begin{enumerate}
\item Fully reflective ring (ring particles have albedo = 1 and ring is opaque): This idealized scenario represents the maximum possible reflectance and is broadly applicable to the visible and near-infrared if the ring is composed primarily of water ice with specific grain sizes. For example, 1~$\mu$m-sized grains can yield an albedo of $\sim$0.95 over the 0.2--2~$\mu$m range (see Fig.4 of \citealt{Clark2019}).
\item Saturn-like ring reflectivity (ring particles have albedo = 0.36 and ring is opaque): Figure \ref{fig:AlbedoAndFilters} shows the reflectance spectrum and albedo of Saturn’s rings overlaid with the transmission curves of JWST/NIRCam wide-band filters. We adopted the reflectivity of Saturn's B ring expected in the F200W filter in this case (albedo = 0.36; \citealt{Ciarniello2019}). 
\end{enumerate}

Under these assumptions, the observed contrast between the planetary ring system and the host star is given by, 

\begin{equation}
\text{contrast} = \ \frac{albedo *\pi (R_{\mathrm{out}}^2 - R_{\mathrm{in}}^2)}{4\pi d^2},
\label{eq:contrast}
\end{equation}

\noindent where $R_{\mathrm{in}}$/$R_{\mathrm{out}}$ are the inner/outer radii of the ring and $d$ is the projected star–to-planet/ring orbital separation.  

When referring to the area of Saturn's rings in this work, we use a reference value corresponding to the inner edge of Saturn’s D ring of $R_{\mathrm{in}}$ = 66,900 km \citep{Murray1999,French2017} and the outer edge of Saturn’s A ring of $R_{\mathrm{out}}$ = 136,780 km \citep{ElMoutamid2016}.  In later sections, we vary the size of the rings discussed by changing the value of the outer ring radius.

\subsection{Limit to the Outer Radius of an Exoring}

Saturn has the largest ring system in our solar system, yet there are virtually no observational constraints on how large an exoring system could be. As a theoretical upper limit, one can assume that a ring system could extend as far as the planet's Hill sphere. The Hill radius ($R_{\mathrm{H}}$) increases with planetary mass ($m_2$), orbital distance ($a$), and decreases with stellar mass ($m_1$), following:

\begin{equation}
R_{\mathrm{H}} \approx a (1-e) \left( \frac{m_2}{3(m_1 + m_2)} \right)^{1/3}.
\end{equation}

For a Sun–Saturn analog system, the Hill radius extends to $R_{\mathrm{out}} = 0.44$ AU ($6.5 \times 10^7$ km), more than 470 times the outer radius of Saturn’s A ring. For the mature super-Jupiter exoplanet Eps Indi Ab, the Hill radius is 2.3 AU ($3.4 \times 10^8$ km), roughly 2515 times larger than Saturn's A ring. 

Although exorings have yet to be detected in these older systems, 
circumplanetary disks observed around young planets exhibit radii comparable to these theoretical maxima. For example, the disk around PDS 70c spans $\sim$1.2 AU which is near its Hill radius \citep{2021ApJ...916L...2B}. The disk surrounding J1407b has an estimated radius of up to 0.6 AU ($9 \times 10^7$ km), approximately 0.15 of its Hill radius (200$\times$ the radius of Saturn’s ring system) \citep{vanWerkhoven2014}.

While it may be possible for a high-opacity distribution of dust or ice grains to extend out to a significant portion of the Hill sphere, the planetary ring systems found around the Solar System giant planets contain regions of high and low optical depth. The most opaque rings lie closer to the Roche limit, the distance within which tidal forces from the planet prevent moons from existing. 
Distances a few times that of the Roche limit may then serve as a more empirically motivated boundary for the outer extent of exoring structures that would be of high enough opacity to be detectable. 
For Eps Indi Ab, the Roche limit is roughly  
$3.5\times10^5$ km (2.6$\times$ Saturn's A-ring radius) 
for a moon that is a fluid-body with the density of water ice.\footnote{The Eps Indi Ab Roche limit calculation was made assuming the density of Eps Indi Ab was $\rho_{planet} = 6074\,\mathrm{kg/m^3}$ and a moon of $\rho_{\text{moon}} = 917\,\mathrm{kg/m^3}$.}

Although we lack strong examples of planetary ring systems extending far beyond the Roche limit in our own Solar System, the diversity of planetary architectures revealed by exoplanet surveys suggests that even extreme configurations should not be dismissed without observational evidence. To date, the transition from young circumplanetary disks to mature ring systems remains largely unexplored and the occurrence rates of such exoring systems have yet to be measured.

\section{Results \label{sec:results}}

\begin{figure}[b]
\centering
\includegraphics[width=0.45\textwidth]{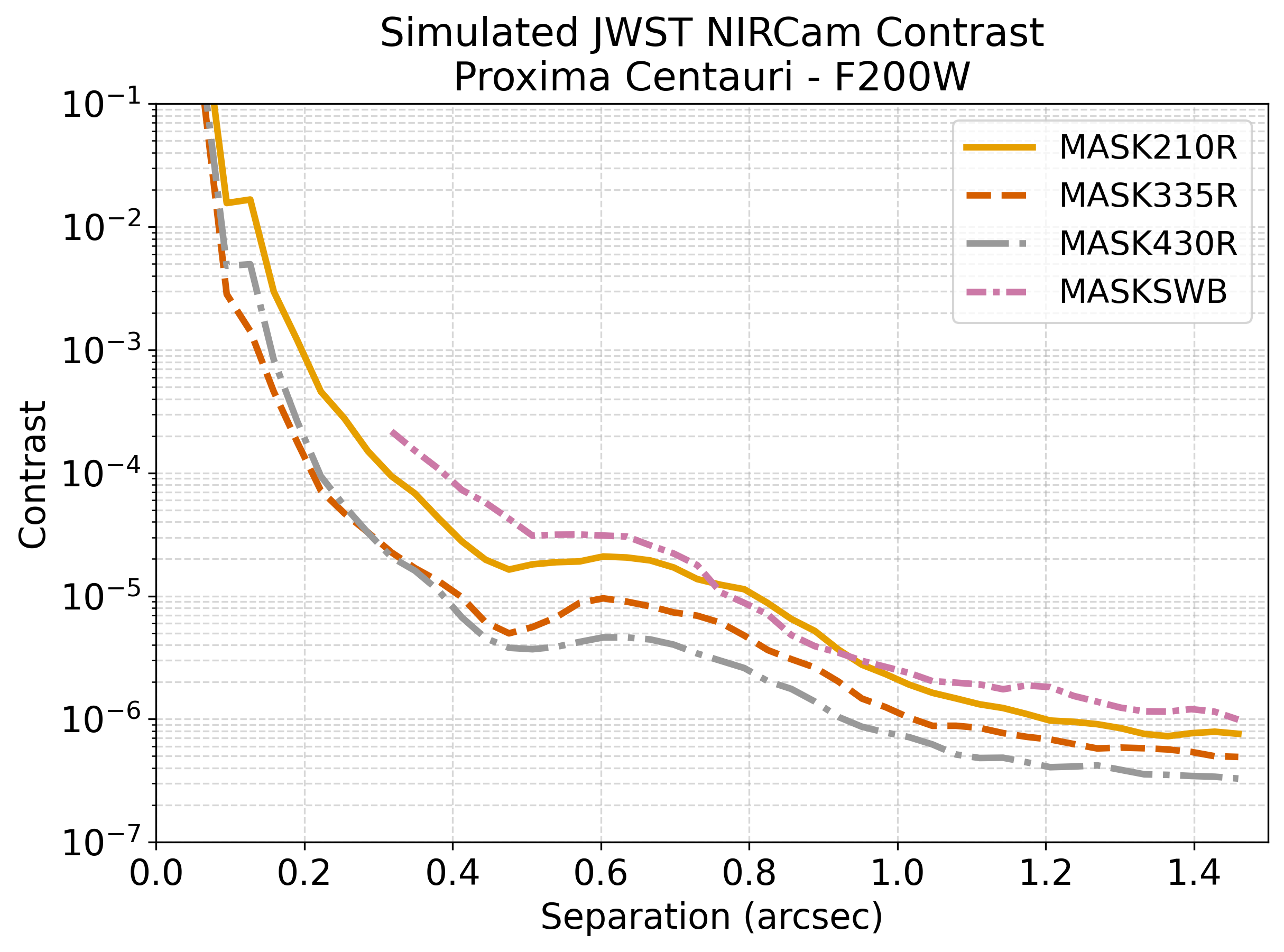}
\caption{\textbf{NIRCam Contrast Simulations To Determine the Optimal Mask}. An example of the \texttt{PanCAKE} contrast simulations are shown for Proxima Centauri with the F200W filter. This example is consistent across both targets and other filters.  MASK335R and MASK430R are the best performing masks in simulation. }
\label{fig:PanCAKE_F200W_ProxCen}
\end{figure}

Figure \ref{fig:PanCAKE_F200W_ProxCen} shows the results of the PanCAKE sims for Proxima Centauri for the F200W filter. We find that the best performing masks for NIRCam's short wavelength filters are MASK335R and MASK430R.   While MASK430R outperforms MASK335R at some separations, we elected to baseline the remainder of the study with MASK335R because of the asymmetry in the field of view of MASK430R at short wavelengths,\footnote{See jdocs for field of view information: \url{https://jwst-docs.stsci.edu/jwst-near-infrared-camera/nircam-observing-modes/nircam-coronagraphic-imaging}} and 
MASK335R is currently the mask of choice for the majority of JWST exoplanet direct imaging programs (see GO 4050 \citep{Carter2023jwst}, GO 5835 \citep{Carter2024jwstprop}, GO 6012 \citep{Millar-Blanchaer2024jwstprop}, GO 6122 \citep{CoolKidsprop}, and SURVEY 6005\footnote{SURVEY 6005: \url{https://www.stsci.edu/jwst/science-execution/program-information?id=6005})}).

\begin{figure*}
\gridline{
  \fig{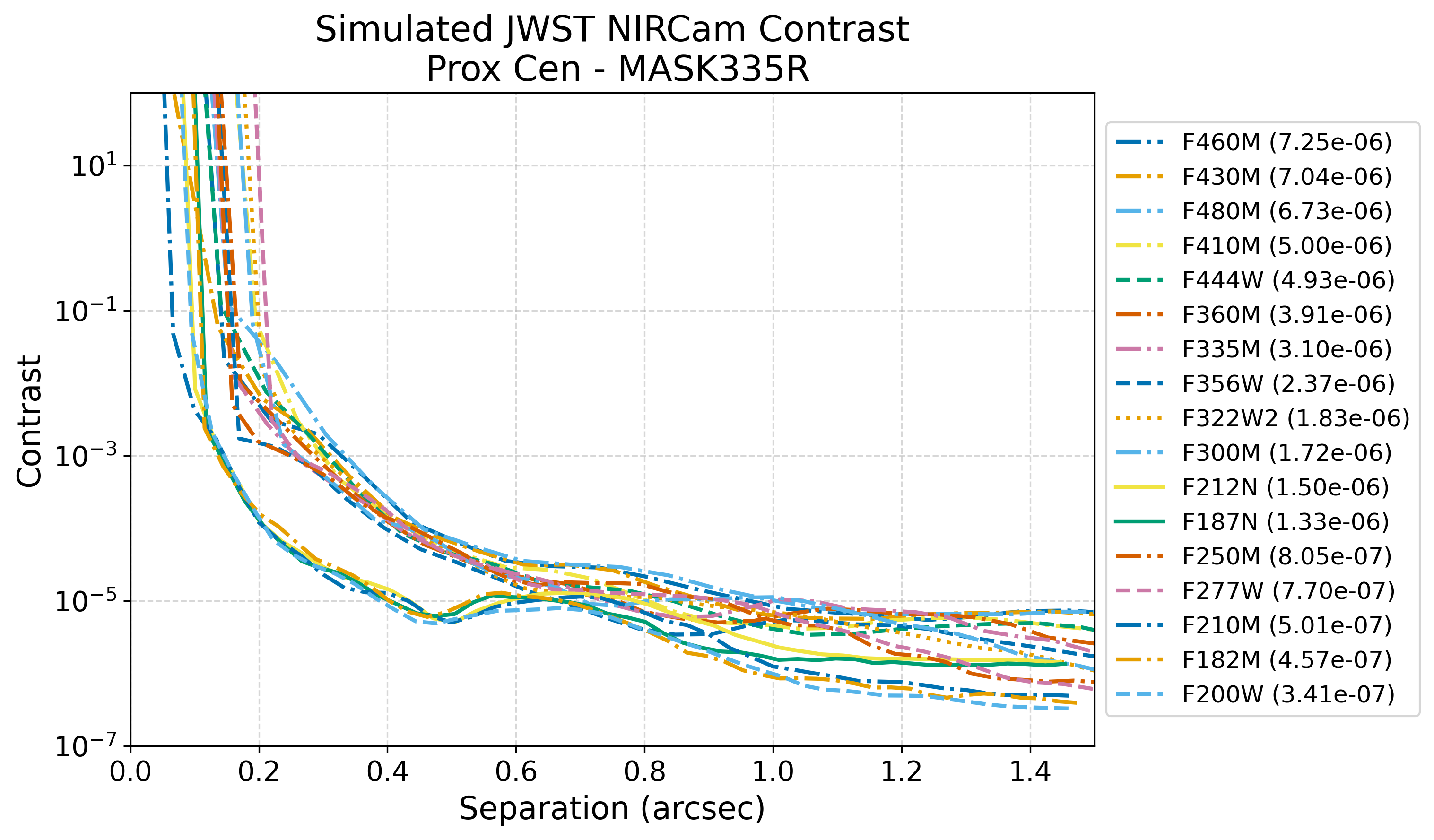}{0.5\textwidth}{(a) Proxima Centauri (MV5.5; 1.3pc)}
  \fig{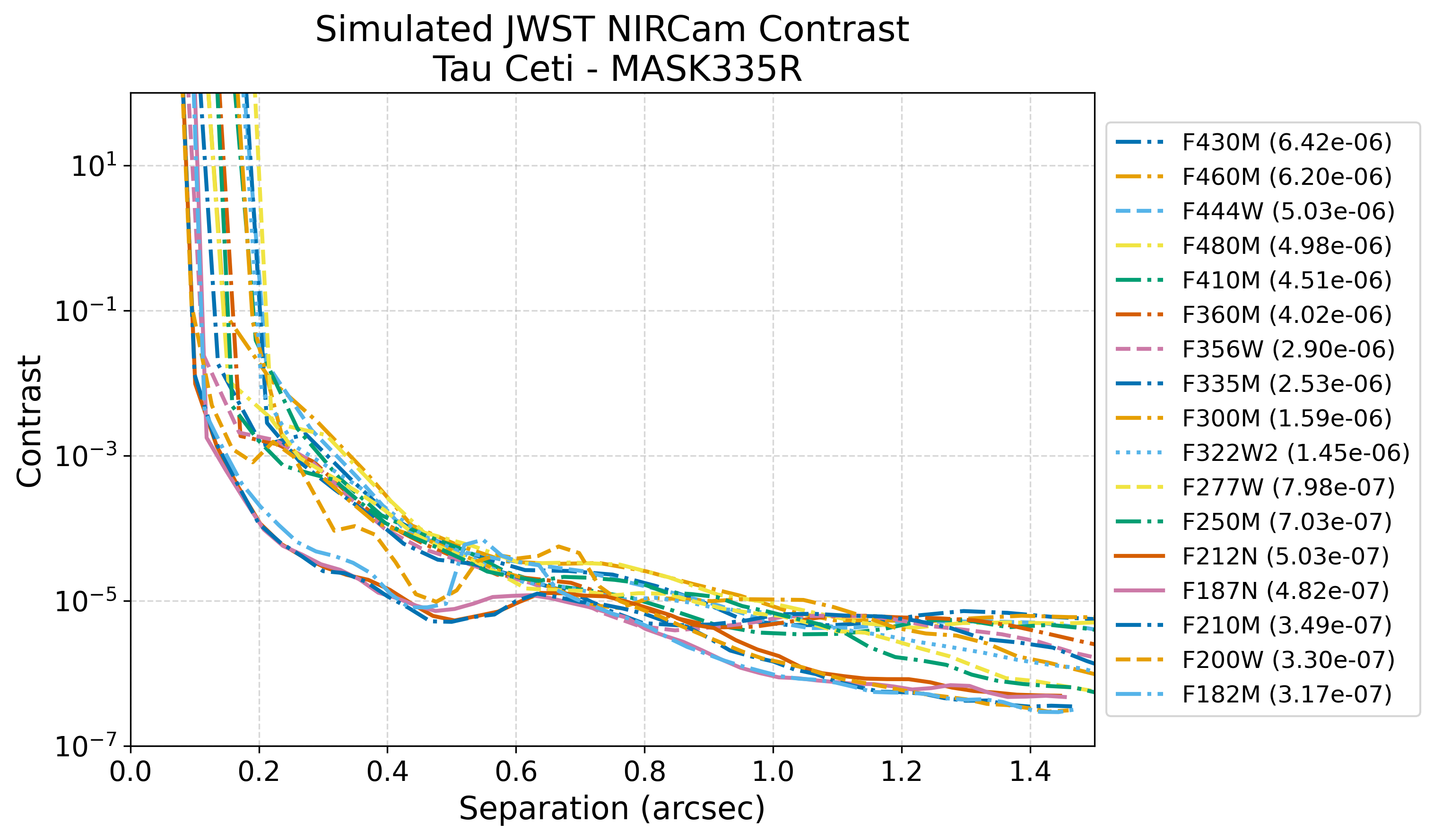}{0.5\textwidth}{(b) Tau Ceti (G8V; 3.65pc)}
}
\caption{\textbf{NIRCam Corongraphy Simulations with MASK335R to Determine Optimal Filter}. \texttt{PanCAKE} was used to simulate the contrast achievable with NIRCam coronagraphy. The contrast predictions for the narrow filters (N) are shown with solid lines, the medium filters (M) are shown in the solid-dot, the wide filters (W) are shown in the dash-dot, and the extra wide (W2) are shown in dotted. The values of the simulated contrast at 1.4 arcsec are listed in the legend. The filters are ranked in the order of least to most sensitive. The three filters yielding the best sensitivity for both targets were: F182M, F200W, and F210M.}
\label{fig:PanCAKE_MASK335R}
\end{figure*}

Figure \ref{fig:PanCAKE_MASK335R} presents the contrast simulations for Proxima Centauri and Tau Ceti for the full set of available filters to perform NIRCam coronagraphy using the MASK335R mask. To determine the optimal filter, we ranked each filter by the contrast value achieved at 1.4 arcsec. This separation is within the limits of \texttt{PanCAKE} for both the short- and long-wavelength filters and is expected to be near the background-limited regime. The contrast  for each filter for Prox Cen and Tau Ceti are listed in the legends of Figure \ref{fig:PanCAKE_MASK335R}.  

The three filters that predicted the best contrast performance for Proxima Centauri and Tau Ceti were F182M, F200W, and F210N. 
The best achievable contrast was $3.4 \times 10^{-7}$ at 1.4" (1.8\,AU) for Proxima Centauri using F200W and $3.2 \times 10^{-7}$ at 1.4" (5.1\,AU) for Tau Ceti using F182N.  
These filters provide the best JWST contrast performance independently of considering where rings may be most reflective. However, these filters happen to be within the wavelength range where Saturn's B-ring albedo remains reasonably high.

\begin{figure*}
\gridline{\fig{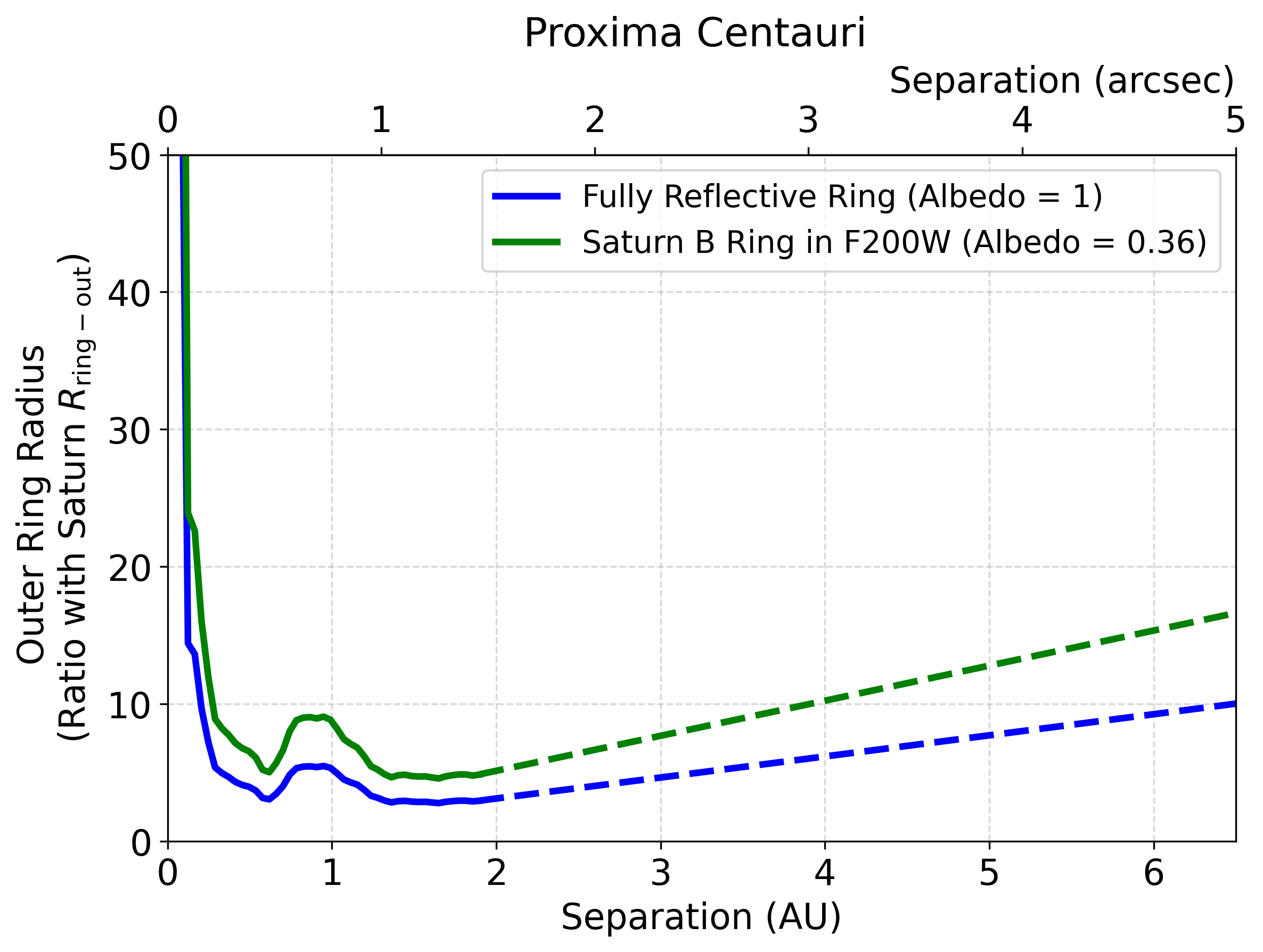}{0.5\textwidth}{(a) Proxima Centauri (MV5.5; 1.3pc)}
\fig{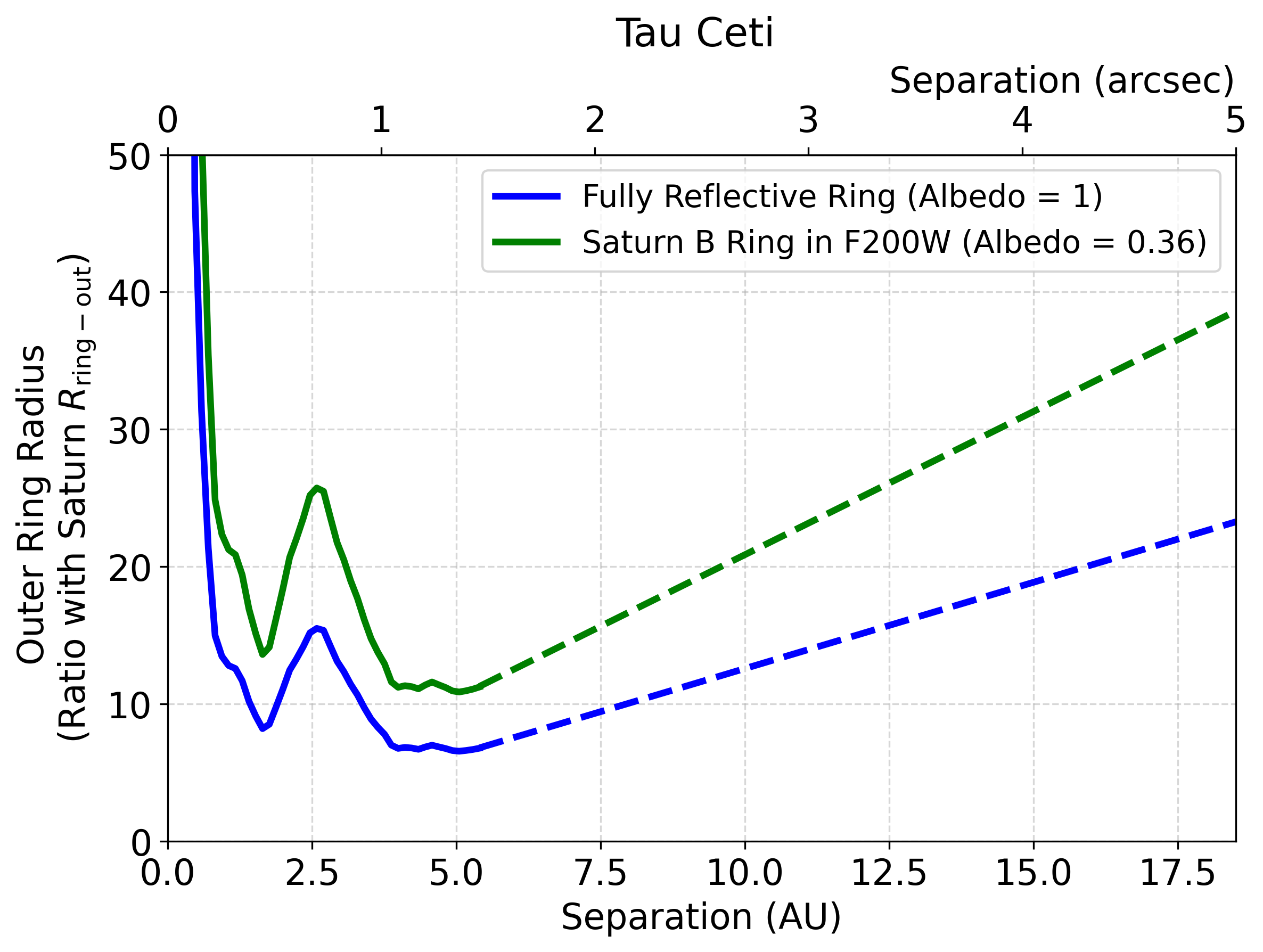}{0.5\textwidth}{(b) Tau Ceti (G8V; 3.65pc)}}
\caption{\textbf{Ring Size Detectable with JWST}. We used the results from the simulations of NIRCam coronagraphy F200W + MASK335R to estimate the smallest ring size detectable for Proxima Centauri and Tau Ceti in the fully reflective case \textit{(blue)} and the case where the albedo matches that of Saturn's rings at F200W \textit{(green)}. 
The estimate is shown to the edge of the SUB320 field of view (5$''$ radius). The dotted lines represent where the contrast was extrapolated linearly from the final 3 points of the \texttt{PanCAKE} simulation. While ring systems smaller than 2.8 the radius of Saturn's rings are not likely to be detectable with JWST through reflected light direct imaging, large ring systems (tens to thousands of times the size of Saturn's rings) surrounding planets in nearby systems could be possible to detect using NIRCam coronagraphy.  }

\label{fig:ringsize}
\end{figure*}

We then converted the simulated F200W contrast curves into estimates of the smallest detectable ring sizes for each system, as shown in Figure \ref{fig:ringsize}. These estimates are expressed in terms of the smallest outer ring radius of the exoring system normalized to the outer radius of Saturn’s A ring ($R_{out}=136,780$\,km). 
For Proxima Centauri, the minimum detectable ring size for the fully-reflective case corresponds to 2.8$\times$ the radius of Saturn’s rings, with peak sensitivity between 1.3 and 1.9 AU (1.0–1.5\,arcsec). If the ring reflectivity instead matches that of Saturn’s rings (albedo of 0.36 instead of 1), the detection threshold increases to 4.6 times Saturn’s ring radius.
In the Tau Ceti system, the minimum detectable ring size under the fully reflective assumption is 6.6 times the radius of Saturn’s rings. Assuming Saturn-ring-like reflectivity, this threshold increases to 10.9 times Saturn’s ring radius with the optimal sensitivity to planets between 4.0 and 5.3 AU (1.1–1.5 arcsec) separation.

Although NIRCam coronagraphy achieves similar contrast performance at a given angular separation in both systems, smaller exorings are detectable around Proxima Centauri than around Tau Ceti due to its closer distance to the observer. Based on Figure~\ref{fig:ringsize}, we find that ring systems with radii up to 10 times that of Saturn’s are potentially detectable around exoplanets at orbital separations of approximately 0.5–10\,AU in most systems within a few parsecs. However, detectability is limited to a subset of ring properties depending on the ring's reflectivity and viewing inclination.
Our results suggest that a planetary ring system with an outer radius of 2.8 times that of Saturn’s rings—corresponding to the Proxima Centauri fully reflective case—represents a practical lower size limit for the detectability of reflected light from an exoring system with JWST NIRCam.

\section{Discussion \label{sec:discussion}}

Simulations of JWST NIRCam coronagraphy indicate that while JWST is capable of detecting reflected light from large exoring systems, it cannot detect a ring system comparable in size to Saturn’s at any orbital separation—even for the nearest stellar systems. This result aligns with expectations, as JWST was not designed for near infrared reflected-light imaging of exoplanetary rings.

Nonetheless, the ability to detect ring systems several times larger than Saturn’s is a somewhat surprising capability. Observers should keep in mind that if an exoplanet candidate is identified within a suitable separation to receive a sufficient proportion of light from its star and showing a bright excess of flux at F200W as compared to F444W, it may be possible that this excess could be explained by a large-ring system with a similar albedo to Saturn's B-ring (which does not reflect much light in F444W).    Many JWST direct imaging programs already include NIRCam F200W coronagraphic imaging as part of their observations, and so it is possible that an accidental detection would be possible for one of these NIRCam surveys. 
Leveraging archival NIRCam F200W data could provide a first step toward constraining the occurrence rate of giant exoplanetary ring systems.  Such an  occurrence rate analysis would require care in considering the many degeneracies that could still leave a large ring systems undetectable to JWST, including misalignment of the rings geometry with respect to the observer,  ring systems being made of a non-reflective material, or rings lacking the sufficient grain density to have an opacity to be strongly reflective.

\begin{figure*}
    \centering
    \includegraphics[width=0.95\linewidth]{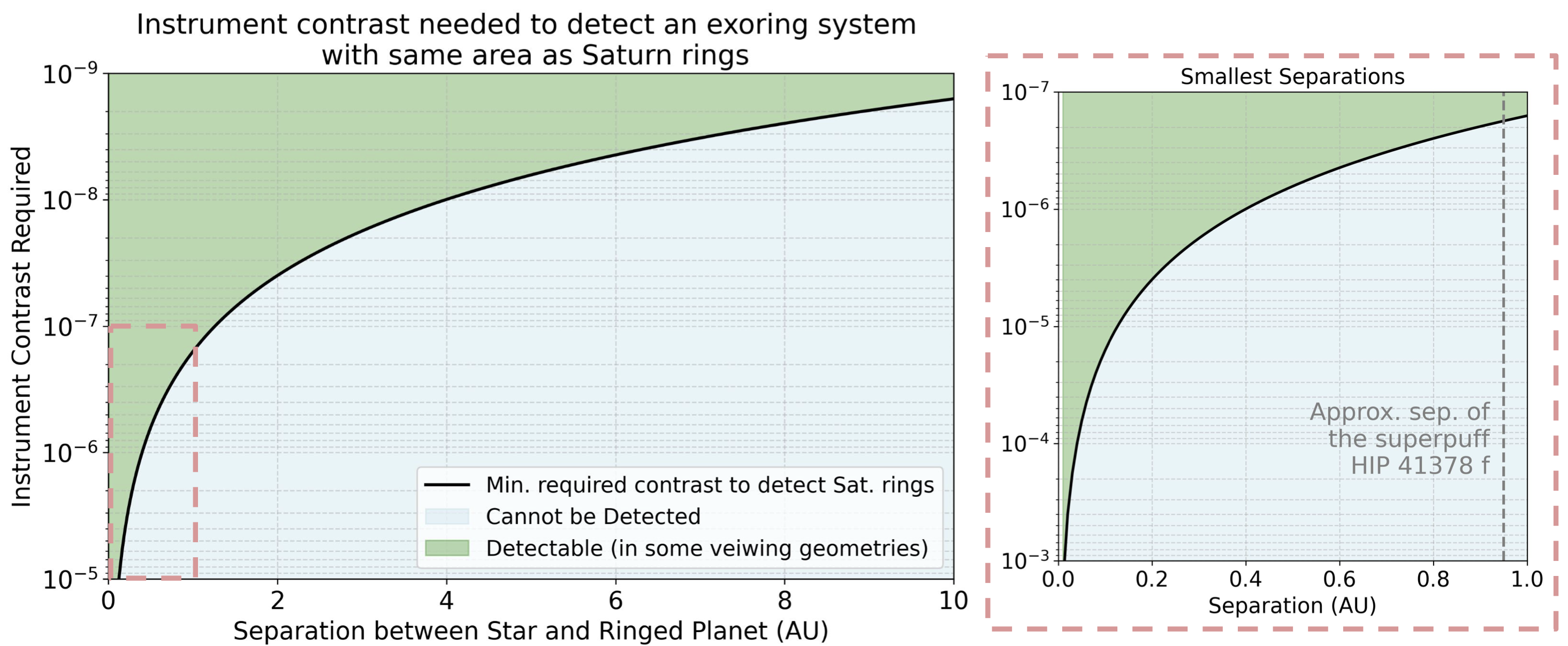}
    \caption{\textbf{Minimum instrument contrast needed to detect an exoring system with same area as Saturn rings}. Instruments capable of reaching contrasts above that of the black line at a given separation \textit{(shaded in green)} would be capable of detecting face-on, fully-reflective exoring systems with the equivalent area of Saturn's rings. As such, the black line provides the most optimistic contrast that would be needed to make the detection. Current ground-based instrumentation may be capable of reaching the contrasts required at tight separations \textit{(see zoomed plot on right)}. The approximate AU separation of the superpuff HIP 41378\,f is plotted in gray to show an example of an exoplanet theorized to have a ring system.\footnote{The separation of HIP 41378 f was calculated using the central values for the period and the stellar mass provided in \cite{Vanderburg2016}. While the radius of this ring system is not fully known, the predictions in \cite{Akinsanmi2020} indicate that this exoring radius would be likely be smaller than that of Saturn's ring system.} However, large exoring systems are predicted to be rare on tighter orbits due to the gravitational influence of the star. 
    While the contrasts to detect a exoring with the same area as Saturn's rings on orbits outside of 1 AU is out of reach for JWST, 
    the contrast performance of future instrumentation is expected to reach this needed sensitivity when directly imaging the nearest neighboring systems. }
    \label{fig:ContrasttofindSaturn}
\end{figure*}

Figure \ref{fig:ContrasttofindSaturn} plots the instrument contrast required to discover an exoring system with the equivalent area of Saturn at varying separations from its host star.  We again adopt the assumptions outlined in Section \ref{sec:methods-rings} with the fully reflective case such that this plot represents the minimal instrument contrast that would be required to discover a ring system with the same area as Saturn's rings.  

To detect an exoring system the same area as Saturn's around an exoplanet at the same orbital separation as Saturn ($a = 9.58$\,AU), an instrument contrast of $1.73\times10^{-9}$ or better would be required.  This contrast requirement is well within the Habitable Worlds Observatory (HWO) prediction and around the target goal for Nancy Grace Roman Space Telescope Coronagraph Instrument (Roman CGI; see Figure 16 in \citealt{Follette2023}). Because the instrument threshold of detecting small exoring systems will be met with future observatories, considerations must be given for how to disentangle the properties of exoplanets from the properties of exorings when interpreting the flux measurements from a detected source. We are currently exploring the detectability and characterization of exorings with these future observatories in detail in forthcoming work.

The instrument contrast requirements for detecting an exoring with the same area as Saturn's rings becomes more relaxed at smaller separations because of the greater amount of light that reaches the planet at close-in separations. For example, at 1 AU an instrumentation contrast of only $1.6\times 10^{-7}$ is required in the fully-reflective, face-on case.  Some ground-based instrumentation may be capable of reaching these contrasts at the required separations when targeting the nearest-neighboring systems. For example, the Gemini Planet Imager was shown to be capable of achieving a contrast of $10^{-7}$ by 0.7 arcsec (equivalent of 0.9 AU if observing Proxima Centauri) \citep{Follette2023}. The Planetary Camera and Spectrograph (PCS) for the Extremely Large Telescope (ELT) is expected to reach $10^{-9}$ at 0.1 arcsec \citep{2021Msngr.182...38K}. 
However, at particularly small orbital separations, large ring systems are not expected to persist due to the reduced size of the planet’s Hill sphere and increased susceptibility to dynamical perturbations from the nearby host star which can destabilize extended ring structures.

Finally, we briefly examined the relative brightness of Saturn’s rings compared to the planet in the mid-infrared (10–25$\mu$m),  \citep{2023Icar..39215347B,2023JGRE..12807924F,2024JGRE..12908236H}. At the wavelengths accessible JWST MIRI, Saturn’s thermal emission dominates over flux reflected by its rings. This suggests that studying exorings around cold giant planets with similar temperture (95K) and atmosphere to Saturn will be more favorable at shorter wavelengths where reflected light is more prominent, such as in the NIRCam bandpasses. This advantage stems from the typically higher albedo of icy ring particles. However, the brightness contrast between the rings and planet can vary depending on the composition of both planet and ring. For example, an exoplanet with highly reflective clouds paired with a ring system contaminated with darker material or composed of silicates rather than water ice would display a different tradespace. Observations of Saturn demonstrate that at continuum wavelengths its atmospheric reflectivity can be quite high \citep[e.g., Fig. 8 of][]{Wang2024}, and detailed spectra of both the planet and its rings reveal strong wavelength-dependent behavior \citep[e.g., Fig. 10.10 in][]{Fletcher2015}. These complexities underscore the importance of both wavelength selection and system composition in planning future exoring searches.

\section{Conclusions \label{sec:conclusions}}

In this work, we explored the potential of JWST NIRCam coronagraphy to detect reflected light from exoplanetary ring systems. Using the \texttt{PanCAKE} Python package, we simulated contrast limits for two nearby, mature star systems. Our key findings are:

\begin{itemize}
\item Among NIRCam coronagraphic modes, the F200W filter paired with the MASK335R provides one of the most favorable configurations for detecting reflected light from exorings.

\item Exoring systems with radii smaller than 2.8 times that of Saturn’s rings are unlikely to be detectable via direct imaging with NIRCam coronagraphy at any orbital separation. Consequently, photometric planet characterization using F200W is unlikely to be significantly affected by ring flux unless the rings are substantially larger than those found in our Solar System.

\item 
Rings $\sim$10$\times$ the size of Saturn’s are potentially detectable at separations of 0.5--10 AU for the closest systems within a few parsecs. This detectability is dependent on the reflectivity of the rings and their viewing geometry.

\item Existing NIRCam F200W coronagraphic observations of nearby stars could be leveraged to constrain the occurrence rate of large exoring systems ($\gtrsim$100$\times$ the size of Saturn’s rings). 
Such constraints represent an important step toward empirically assessing the prevalence of giant exorings around wide-orbit, mature exoplanets.

\end{itemize}

These results demonstrate both the current limitations and potential opportunities for using JWST coronagraphy to detect and constrain the presence of large exoring systems around mature exoplanets. 
As the JWST NIRCam coronagraphy archival data continues to grow, systematic searches for large exorings may provide an unexpected legacy on the architectures of planetary systems around nearby stars.

\begin{acknowledgments}

The authors would like to acknowledge the thousands of people who dedicated themselves to the design, construction, and commissioning of JWST and the current staff at Space Telescope Sciences who operate the telescope. The authors would also like to thank Aarynn Carter and Marshall Perrin for their role in developing and maintaining \texttt{PanCAKE}.

\end{acknowledgments}

\begin{contribution}

This paper was inspired by the ongoing work by Sam Hopper and Mary Anne Limbach to explore the detectability of exoring systems for HWO and Roman CGI.  Sam Hopper's work assembling the literature measurements of the rings of Saturn and considering their observability as an exoring analog led to the natural next questions that were the foundation of this work. 
Rachel Bowens-Rubin completed the analysis and writing for Sections 2 - 5 as well as edits to the abstract and Section 1. Mary Anne Limbach authored the initial draft of the abstract and Section 1 and provided edits on the rest of the manuscript. Klaus Subbotina Stephenson provided essential support and guidance for setting up and running the \texttt{PanCAKE} simulations. Matson Garza supported the calculations of the Roche limits and Hill sphere values in Section 2.3. Leigh Fletcher and Matthew Hedman provided their expertise regarding planetary ring systems within the solar system to make recommendations and suggestions in the review of early drafts of the manuscript. 

\end{contribution}


\software{\texttt{PanCAKE} \citep{Girard2018, Perrin2018, CarterPancake}, ChatGPT was used to improve the wording at the sentence level and to assist in coding.}

\bibliography{sample7}{}

\begin{thebibliography}{}
\expandafter\ifx\csname natexlab\endcsname\relax\def\natexlab#1{#1}\fi
\providecommand{\url}[1]{\href{#1}{#1}}
\providecommand{\dodoi}[1]{doi:~\href{http://doi.org/#1}{\nolinkurl{#1}}}
\providecommand{\doeprint}[1]{\href{http://ascl.net/#1}{\nolinkurl{http://ascl.net/#1}}}
\providecommand{\doarXiv}[1]{\href{https://arxiv.org/abs/#1}{\nolinkurl{https://arxiv.org/abs/#1}}}

\bibitem[{B. {Akinsanmi} {et~al.}(2020){Akinsanmi}, {Santos}, {Faria}, {Oshagh}, {Barros}, {Santerne}, \& {Charnoz}}]{Akinsanmi2020}
{Akinsanmi}, B., {Santos}, N.~C., {Faria}, J.~P., {et~al.} 2020, \bibinfo{title}{{Can planetary rings explain the extremely low density of HIP 41378 f?},} \aap, 635, L8, \dodoi{10.1051/0004-6361/202037618}

\bibitem[{G. {Anglada-Escud{\'e}} {et~al.}(2016){Anglada-Escud{\'e}}, {Amado}, {Barnes}, {Berdi{\~n}as}, {Butler}, {Coleman}, {de La Cueva}, {Dreizler}, {Endl}, {Giesers}, {Jeffers}, {Jenkins}, {Jones}, {Kiraga}, {K{\"u}rster}, {L{\'o}pez-Gonz{\'a}lez}, {Marvin}, {Morales}, {Morin}, {Nelson}, {Ortiz}, {Ofir}, {Paardekooper}, {Reiners}, {Rodr{\'\i}guez}, {Rodr{\'\i}guez-L{\'o}pez}, {Sarmiento}, {Strachan}, {Tsapras}, {Tuomi}, \& {Zechmeister}}]{Anglada-Escudé2016}
{Anglada-Escud{\'e}}, G., {Amado}, P.~J., {Barnes}, J., {et~al.} 2016, \bibinfo{title}{{A terrestrial planet candidate in a temperate orbit around Proxima Centauri},} \nat, 536, 437, \dodoi{10.1038/nature19106}

\bibitem[{L. {Arnold} \& J. {Schneider}(2006){Arnold} \& {Schneider}}]{ArnoldandSchneider2006}
{Arnold}, L., \& {Schneider}, J. 2006, in IAU Colloq. 200: Direct Imaging of Exoplanets: Science \& Techniques, ed. C.~{Aime} \& F.~{Vakili}, 105--110, \dodoi{10.1017/S1743921306009173}

\bibitem[{J. {Bae} {et~al.}(2022){Bae}, {Teague}, {Andrews}, {Benisty}, {Facchini}, {Galloway-Sprietsma}, {Loomis}, {Aikawa}, {Alarc{\'o}n}, {Bergin}, {Bergner}, {Booth}, {Cataldi}, {Cleeves}, {Czekala}, {Guzm{\'a}n}, {Huang}, {Ilee}, {Kurtovic}, {Law}, {Le Gal}, {Liu}, {Long}, {M{\'e}nard}, {{\"O}berg}, {P{\'e}rez}, {Qi}, {Schwarz}, {Sierra}, {Walsh}, {Wilner}, \& {Zhang}}]{2022ApJ...934L..20B}
{Bae}, J., {Teague}, R., {Andrews}, S.~M., {et~al.} 2022, \bibinfo{title}{{Molecules with ALMA at Planet-forming Scales (MAPS): A Circumplanetary Disk Candidate in Molecular-line Emission in the AS 209 Disk},} \apjl, 934, L20, \dodoi{10.3847/2041-8213/ac7fa3}

\bibitem[{J.~W. {Barnes} \& J.~J. {Fortney}(2004){Barnes} \& {Fortney}}]{BarnesandFortney2004}
{Barnes}, J.~W., \& {Fortney}, J.~J. 2004, \bibinfo{title}{{Transit Detectability of Ring Systems around Extrasolar Giant Planets},} \apj, 616, 1193, \dodoi{10.1086/425067}

\bibitem[{C.~A. {Beichman} {et~al.}(2017){Beichman}, {Rieke}, {Bouwman}, {Gaspar}, {Leisenring}, {Su}, \& {Ygouf}}]{GO1193prop}
{Beichman}, C.~A., {Rieke}, G., {Bouwman}, J., {et~al.} 2017, \bibinfo{title}{{Coronagraphic Imaging of Young Planets and Debris Disk with NIRCam and MIRI},}, JWST Proposal. Cycle 1, ID. \#1193

\bibitem[{M. {Benisty} {et~al.}(2021){Benisty}, {Bae}, {Facchini}, {Keppler}, {Teague}, {Isella}, {Kurtovic}, {P{\'e}rez}, {Sierra}, {Andrews}, {Carpenter}, {Czekala}, {Dominik}, {Henning}, {Menard}, {Pinilla}, \& {Zurlo}}]{2021ApJ...916L...2B}
{Benisty}, M., {Bae}, J., {Facchini}, S., {et~al.} 2021, \bibinfo{title}{{A Circumplanetary Disk around PDS70c},} \apjl, 916, L2, \dodoi{10.3847/2041-8213/ac0f83}

\bibitem[{M.~S. {Bessell}(1991){Bessell}}]{Bessell1991}
{Bessell}, M.~S. 1991, \bibinfo{title}{{The Late M Dwarfs},} \aj, 101, 662, \dodoi{10.1086/115714}

\bibitem[{J.~S.~D. {Blake} {et~al.}(2023){Blake}, {Fletcher}, {Orton}, {Antu{\~n}ano}, {Roman}, {Kasaba}, {Fujiyoshi}, {Melin}, {Bardet}, {Sinclair}, \& {Es-Sayeh}}]{2023Icar..39215347B}
{Blake}, J. S.~D., {Fletcher}, L.~N., {Orton}, G.~S., {et~al.} 2023, \bibinfo{title}{{Saturn's seasonal variability from four decades of ground-based mid-infrared observations},} \icarus, 392, 115347, \dodoi{10.1016/j.icarus.2022.115347}

\bibitem[{W.~F. {Bottke} {et~al.}(2005){Bottke}, {Durda}, {Nesvorn{\'y}}, {Jedicke}, {Morbidelli}, {Vokrouhlick{\'y}}, \& {Levison}}]{2005Icar..175..111B}
{Bottke}, W.~F., {Durda}, D.~D., {Nesvorn{\'y}}, D., {et~al.} 2005, \bibinfo{title}{{The fossilized size distribution of the main asteroid belt},} \icarus, 175, 111, \dodoi{10.1016/j.icarus.2004.10.026}

\bibitem[{R. {Bowens-Rubin} {et~al.}(2024){Bowens-Rubin}, {Limbach}, {Carter}, {Ertel}, {Girard}, {Hinz}, {Matthews}, {Morley}, {Mukherjee}, {Salama}, {Vanderburg}, \& {Wagner}}]{CoolKidsprop}
{Bowens-Rubin}, R., {Limbach}, M.~A., {Carter}, A., {et~al.} 2024, \bibinfo{title}{{Cool kids on the block: The direct detection of cold ice giants and gas giants orbiting young low-mass neighbors},}, JWST Proposal. Cycle 3, ID. \#6122

\bibitem[{R. {Bowens-Rubin} {et~al.}(2025){Bowens-Rubin}, {Mang}, {Limbach}, {Carter}, {Stevenson}, {Wagner}, {Strampelli}, {Morley}, {Kennedy}, {Matthews}, {Vanderburg}, \& {Salama}}]{Bowens-Rubin2025}
{Bowens-Rubin}, R., {Mang}, J., {Limbach}, M.~A., {et~al.} 2025, \bibinfo{title}{{NIRCam Yells at Cloud: JWST MIRI Imaging Can Directly Detect Exoplanets of the Same Temperature, Mass, Age, and Orbital Separation as Saturn and Jupiter},} \apjl, 986, L26, \dodoi{10.3847/2041-8213/addbde}

\bibitem[{R.~M. {Canup}(2010){Canup}}]{2010Natur.468..943C}
{Canup}, R.~M. 2010, \bibinfo{title}{{Origin of Saturn's rings and inner moons by mass removal from a lost Titan-sized satellite},} \nat, 468, 943, \dodoi{10.1038/nature09661}

\bibitem[{R.~M. {Canup} \& W.~R. {Ward}(2002){Canup} \& {Ward}}]{2002AJ....124.3404C}
{Canup}, R.~M., \& {Ward}, W.~R. 2002, \bibinfo{title}{{Formation of the Galilean Satellites: Conditions of Accretion},} \aj, 124, 3404, \dodoi{10.1086/344684}

\bibitem[{A. {Carter} {et~al.}(2023){Carter}, {Balmer}, {Biller}, {Bogat}, {Bonavita}, {Bowler}, {Calissendorff}, {Fontanive}, {Franson}, {Gagne}, {Girard}, {Hinkley}, {Hoch}, {Kammerer}, {Kennedy}, {Leisenring}, {Macintosh}, {Matthews}, {Meyer}, {Millar-Blanchaer}, {Morley}, {Perrin}, {Pueyo}, {Ray}, {Rebollido}, {Rickman}, {Skemer}, \& {Wang}}]{Carter2023jwst}
{Carter}, A., {Balmer}, W., {Biller}, B., {et~al.} 2023, \bibinfo{title}{{Uncharted Worlds: Towards a Legacy of Direct Imaging of Sub-Jupiter Mass Exoplanets},}, JWST Proposal. Cycle 2, ID. \#4050

\bibitem[{A. {Carter} {et~al.}(2024){Carter}, {Absil}, {Balmer}, {Biller}, {Bogat}, {Bonavita}, {Booth}, {Bowens-Rubin}, {Bowler}, {Calissendorff}, {Chauvin}, {Fontanive}, {Franson}, {Gagne}, {Girard}, {Hinkley}, {Hoch}, {Kammerer}, {Kennedy}, {Leisenring}, {Li}, {Limbach}, {Liu}, {Macintosh}, {Matthews}, {Meyer}, {Millar-Blanchaer}, {Morley}, {Perrin}, {Pueyo}, {Ray}, {Rebollido}, {Rickman}, {Skemer}, {Wang}, {Ward-Duong}, \& {Whiteford}}]{Carter2024jwstprop}
{Carter}, A., {Absil}, O., {Balmer}, W., {et~al.} 2024, \bibinfo{title}{{Into The Spotlight: Unveiling Wide-Separation Sub-Jupiters for Future JWST Characterization},}, JWST Proposal. Cycle 3, ID. \#5835

\bibitem[{A.~L. {Carter} {et~al.}(2021){Carter}, {Skemer}, {Danielski}, {Leisenring}, {Wang}, {Van Gorkom}, {York}, {Adams}, {Biller}, {Girard}, {Hinkley}, {Nickson}, {Perrin}, \& {Pueyo}}]{CarterPancake}
{Carter}, A.~L., {Skemer}, A. J.~I., {Danielski}, C., {et~al.} 2021, in Society of Photo-Optical Instrumentation Engineers (SPIE) Conference Series, Vol. 11823, Techniques and Instrumentation for Detection of Exoplanets X, ed. S.~B. {Shaklan} \& G.~J. {Ruane}, 118230H, \dodoi{10.1117/12.2594501}

\bibitem[{A.~L. {Carter} {et~al.}(2023){Carter}, {Hinkley}, {Kammerer}, {Skemer}, {Biller}, {Leisenring}, {Millar-Blanchaer}, {Petrus}, {Stone}, {Ward-Duong}, {Wang}, {Girard}, {Hines}, {Perrin}, {Pueyo}, {Balmer}, {Bonavita}, {Bonnefoy}, {Chauvin}, {Choquet}, {Christiaens}, {Danielski}, {Kennedy}, {Matthews}, {Miles}, {Patapis}, {Ray}, {Rickman}, {Sallum}, {Stapelfeldt}, {Whiteford}, {Zhou}, {Absil}, {Boccaletti}, {Booth}, {Bowler}, {Chen}, {Currie}, {Fortney}, {Grady}, {Greebaum}, {Henning}, {Hoch}, {Janson}, {Kalas}, {Kenworthy}, {Kervella}, {Kraus}, {Lagage}, {Liu}, {Macintosh}, {Marino}, {Marley}, {Marois}, {Matthews}, {Mawet}, {McElwain}, {Metchev}, {Meyer}, {Molliere}, {Moran}, {Morley}, {Mukherjee}, {Pantin}, {Quirrenbach}, {Rebollido}, {Ren}, {Schneider}, {Vasist}, {Worthen}, {Wyatt}, {Briesemeister}, {Bryan}, {Calissendorff}, {Cantalloube}, {Cugno}, {De Furio}, {Dupuy}, {Factor}, {Faherty}, {Fitzgerald}, {Franson}, {Gonzales}, {Hood}, {Howe}, {Kuzuhara}, {Lagrange}, {Lawson}, {Lazzoni}, {Lew},
  {Liu}, {Llop-Sayson}, {Lloyd}, {Martinez}, {Mazoyer}, {Palma-Bifani}, {Quanz}, {Redai}, {Samland}, {Schlieder}, {Tamura}, {Tan}, {Uyama}, {Vigan}, {Vos}, {Wagner}, {Wolff}, {Ygouf}, {Zhang}, {Zhang}, \& {Zhang}}]{Carter2023}
{Carter}, A.~L., {Hinkley}, S., {Kammerer}, J., {et~al.} 2023, \bibinfo{title}{{The JWST Early Release Science Program for Direct Observations of Exoplanetary Systems I: High-contrast Imaging of the Exoplanet HIP 65426 b from 2 to 16 {\ensuremath{\mu}}m},} \apjl, 951, L20, \dodoi{10.3847/2041-8213/acd93e}

\bibitem[{M. {Ciarniello} {et~al.}(2019){Ciarniello}, {Filacchione}, {D'Aversa}, {Capaccioni}, {Nicholson}, {Cuzzi}, {Clark}, {Hedman}, {Dalle Ore}, {Cerroni}, {Plainaki}, \& {Spilker}}]{Ciarniello2019}
{Ciarniello}, M., {Filacchione}, G., {D'Aversa}, E., {et~al.} 2019, \bibinfo{title}{{Cassini-VIMS observations of Saturn's main rings: II. A spectrophotometric study by means of Monte Carlo ray-tracing and Hapke's theory},} \icarus, 317, 242, \dodoi{10.1016/j.icarus.2018.07.010}

\bibitem[{R.~N. {Clark} {et~al.}(2019){Clark}, {Brown}, {Cruikshank}, \& {Swayze}}]{Clark2019}
{Clark}, R.~N., {Brown}, R.~H., {Cruikshank}, D.~P., \& {Swayze}, G.~A. 2019, \bibinfo{title}{{Isotopic ratios of Saturn's rings and satellites: Implications for the origin of water and Phoebe},} \icarus, 321, 791, \dodoi{10.1016/j.icarus.2018.11.029}

\bibitem[{M. {Damasso} {et~al.}(2020){Damasso}, {Del Sordo}, {Anglada-Escud{\'e}}, {Giacobbe}, {Sozzetti}, {Morbidelli}, {Pojmanski}, {Barbato}, {Butler}, {Jones}, {Hambsch}, {Jenkins}, {L{\'o}pez-Gonz{\'a}lez}, {Morales}, {Pe{\~n}a Rojas}, {Rodr{\'\i}guez-L{\'o}pez}, {Rodr{\'\i}guez}, {Amado}, {Anglada}, {Feng}, \& {G{\'o}mez}}]{Damasso2020}
{Damasso}, M., {Del Sordo}, F., {Anglada-Escud{\'e}}, G., {et~al.} 2020, \bibinfo{title}{{A low-mass planet candidate orbiting Proxima Centauri at a distance of 1.5 AU},} Science Advances, 6, eaax7467, \dodoi{10.1126/sciadv.aax7467}

\bibitem[{X. {Dumusque} {et~al.}(2012){Dumusque}, {Pepe}, {Lovis}, {S{\'e}gransan}, {Sahlmann}, {Benz}, {Bouchy}, {Mayor}, {Queloz}, {Santos}, \& {Udry}}]{Dumusque2012}
{Dumusque}, X., {Pepe}, F., {Lovis}, C., {et~al.} 2012, \bibinfo{title}{{An Earth-mass planet orbiting {\ensuremath{\alpha}} Centauri B},} \nat, 491, 207, \dodoi{10.1038/nature11572}

\bibitem[{M. {El Moutamid} {et~al.}(2016){El Moutamid}, {Nicholson}, {French}, {Tiscareno}, {Murray}, {Evans}, {French}, {Hedman}, \& {Burns}}]{ElMoutamid2016}
{El Moutamid}, M., {Nicholson}, P.~D., {French}, R.~G., {et~al.} 2016, \bibinfo{title}{{How Janus' orbital swap affects the edge of Saturn's A ring?},} \icarus, 279, 125, \dodoi{10.1016/j.icarus.2015.10.025}

\bibitem[{J.~L. {Elliot} {et~al.}(1977){Elliot}, {Dunham}, \& {Mink}}]{1977Natur.267..328E}
{Elliot}, J.~L., {Dunham}, E., \& {Mink}, D. 1977, \bibinfo{title}{{The rings of Uranus},} \nat, 267, 328, \dodoi{10.1038/267328a0}

\bibitem[{J.~P. {Faria} {et~al.}(2022){Faria}, {Su{\'a}rez Mascare{\~n}o}, {Figueira}, {Silva}, {Damasso}, {Demangeon}, {Pepe}, {Santos}, {Rebolo}, {Cristiani}, {Adibekyan}, {Alibert}, {Allart}, {Barros}, {Cabral}, {D'Odorico}, {Di Marcantonio}, {Dumusque}, {Ehrenreich}, {Gonz{\'a}lez Hern{\'a}ndez}, {Hara}, {Lillo-Box}, {Lo Curto}, {Lovis}, {Martins}, {M{\'e}gevand}, {Mehner}, {Micela}, {Molaro}, {Nunes}, {Pall{\'e}}, {Poretti}, {Sousa}, {Sozzetti}, {Tabernero}, {Udry}, \& {Zapatero Osorio}}]{Faria2022}
{Faria}, J.~P., {Su{\'a}rez Mascare{\~n}o}, A., {Figueira}, P., {et~al.} 2022, \bibinfo{title}{{A candidate short-period sub-Earth orbiting Proxima Centauri},} \aap, 658, A115, \dodoi{10.1051/0004-6361/202142337}

\bibitem[{F. {Feng} {et~al.}(2017){Feng}, {Tuomi}, {Jones}, {Barnes}, {Anglada-Escud{\'e}}, {Vogt}, \& {Butler}}]{Feng2017}
{Feng}, F., {Tuomi}, M., {Jones}, H.~R.~A., {et~al.} 2017, \bibinfo{title}{{Color Difference Makes a Difference: Four Planet Candidates around {\ensuremath{\tau}} Ceti},} \aj, 154, 135, \dodoi{10.3847/1538-3881/aa83b4}

\bibitem[{L.~N. {Fletcher} {et~al.}(2015){Fletcher}, {Greathouse}, {Moses}, {Guerlet}, \& {West}}]{Fletcher2015}
{Fletcher}, L.~N., {Greathouse}, T.~K., {Moses}, J.~I., {Guerlet}, S., \& {West}, R.~A. 2015, \bibinfo{title}{{Saturn's Seasonally Changing Atmosphere: Thermal Structure, Composition and Aerosols},} arXiv e-prints, arXiv:1510.05690, \dodoi{10.48550/arXiv.1510.05690}

\bibitem[{L.~N. {Fletcher} {et~al.}(2018){Fletcher}, {Melin}, {Adriani}, {Simon}, {Sanchez-Lavega}, {Donnelly}, {Antu{\~n}ano}, {Orton}, {Hueso}, {Kraaikamp}, {Wong}, {Barnett}, {Moriconi}, {Altieri}, \& {Sindoni}}]{Fletcher2018}
{Fletcher}, L.~N., {Melin}, H., {Adriani}, A., {et~al.} 2018, \bibinfo{title}{{Jupiter{\textquoteright}s Mesoscale Waves Observed at 5 {\ensuremath{\mu}}m by Ground-based Observations and Juno JIRAM},} \aj, 156, 67, \dodoi{10.3847/1538-3881/aace02}

\bibitem[{L.~N. {Fletcher} {et~al.}(2023){Fletcher}, {King}, {Harkett}, {Hammel}, {Roman}, {Melin}, {Hedman}, {Moses}, {Guerlet}, {Milam}, \& {Tiscareno}}]{2023JGRE..12807924F}
{Fletcher}, L.~N., {King}, O. R.~T., {Harkett}, J., {et~al.} 2023, \bibinfo{title}{{Saturn's Atmosphere in Northern Summer Revealed by JWST/MIRI},} Journal of Geophysical Research (Planets), 128, e2023JE007924, \dodoi{10.1029/2023JE007924}

\bibitem[{K.~B. {Follette}(2023){Follette}}]{Follette2023}
{Follette}, K.~B. 2023, \bibinfo{title}{{An Introduction to High Contrast Differential Imaging of Exoplanets and Disks},} \pasp, 135, 093001, \dodoi{10.1088/1538-3873/aceb31}

\bibitem[{R.~G. {French} {et~al.}(2017){French}, {McGhee-French}, {Lonergan}, {Sepersky}, {Jacobson}, {Nicholson}, {Hedman}, {Marouf}, \& {Colwell}}]{French2017}
{French}, R.~G., {McGhee-French}, C.~A., {Lonergan}, K., {et~al.} 2017, \bibinfo{title}{{Noncircular features in Saturn's rings IV: Absolute radius scale and Saturn's pole direction},} \icarus, 290, 14, \dodoi{10.1016/j.icarus.2017.02.007}

\bibitem[{ {Gaia Collaboration}(2020){Gaia Collaboration}}]{GaiaEDR3}
{Gaia Collaboration}. 2020, \bibinfo{title}{{VizieR Online Data Catalog: Gaia EDR3 (Gaia Collaboration, 2020)},}, VizieR On-line Data Catalog: I/350. Originally published in: 2021A\&A...649A...1G \dodoi{10.26093/cds/vizier.1350}

\bibitem[{J.~H. {Girard} {et~al.}(2018){Girard}, {Blair}, {Brooks}, {Brooks}, {Brown}, {Bushouse}, {Canipe}, {Chen}, {Correnti}, {Hagan}, {Hilbert}, {Hines}, {Leisenring}, {Long}, {Nickson}, {Perrin}, {Pontoppidan}, {Pueyo}, {Rajan}, {Riedel}, {Soummer}, {Stansberry}, {Stark}, {Van Gorkom}, \& {York}}]{Girard2018}
{Girard}, J.~H., {Blair}, W., {Brooks}, B., {et~al.} 2018, in Society of Photo-Optical Instrumentation Engineers (SPIE) Conference Series, Vol. 10698, Space Telescopes and Instrumentation 2018: Optical, Infrared, and Millimeter Wave, ed. M.~{Lystrup}, H.~A. {MacEwen}, G.~G. {Fazio}, N.~{Batalha}, N.~{Siegler}, \& E.~C. {Tong}, 106983V, \dodoi{10.1117/12.2314198}

\bibitem[{J.~H. {Girard} {et~al.}(2022){Girard}, {Leisenring}, {Kammerer}, {Gennaro}, {Rieke}, {Stansberry}, {Rest}, {Egami}, {Sunnquist}, {Boyer}, {Canipe}, {Correnti}, {Hilbert}, {Perrin}, {Pueyo}, {Soummer}, {Allen}, {Bushouse}, {Aguilar}, {Brooks}, {Coe}, {DiFelice}, {Golimowski}, {Hartig}, {Hines}, {Koekemoer}, {Nickson}, {Nikolov}, {Kozhurina-Platais}, {Pirzkal}, {Robberto}, {Sivaramakrishnan}, {Sohn}, {Telfer}, {Wu}, {Beatty}, {Florian}, {Hainline}, {Kelly}, {Misselt}, {Schlawin}, {Sun}, {Williams}, {Willmer}, {Stark}, {Ygouf}, {Carter}, {Beichman}, {Greene}, {Roellig}, {Krist}, {Adams Redai}, {Wang}, {Clark}, {Lewis}, \& {Ferry}}]{Girard2022}
{Girard}, J.~H., {Leisenring}, J., {Kammerer}, J., {et~al.} 2022, in Society of Photo-Optical Instrumentation Engineers (SPIE) Conference Series, Vol. 12180, Space Telescopes and Instrumentation 2022: Optical, Infrared, and Millimeter Wave, ed. L.~E. {Coyle}, S.~{Matsuura}, \& M.~D. {Perrin}, 121803Q, \dodoi{10.1117/12.2629636}

\bibitem[{M.~M. {Hedman} {et~al.}(2024){Hedman}, {Tiscareno}, {Showalter}, {Fletcher}, {King}, {Harkett}, {Roman}, {Rowe-Gurney}, {Hammel}, {Milam}, {El Moutamid}, {Cartwright}, {de Pater}, \& {Molter}}]{2024JGRE..12908236H}
{Hedman}, M.~M., {Tiscareno}, M.~S., {Showalter}, M.~R., {et~al.} 2024, \bibinfo{title}{{Water-Ice Dominated Spectra of Saturn's Rings and Small Moons From JWST},} Journal of Geophysical Research (Planets), 129, e2023JE008236, \dodoi{10.1029/2023JE008236}

\bibitem[{M.~M. {Hedman} {et~al.}(2025){Hedman}, {de Pater}, {Cartwright}, {El Moutamid}, {DeColibus}, {Showlater}, {Tiscareno}, {Rowe-Gurney}, {Roman}, {Fletcher}, \& {Hammel}}]{Hedman2025}
{Hedman}, M.~M., {de Pater}, I., {Cartwright}, R., {et~al.} 2025, \bibinfo{title}{{Spectral trends across the rings and inner moons of Uranus and Neptune from JWST NIRCam images},} arXiv e-prints, arXiv:2506.18650, \dodoi{10.48550/arXiv.2506.18650}

\bibitem[{A.~J. {Hesselbrock} \& D.~A. {Minton}(2019){Hesselbrock} \& {Minton}}]{2019AJ....157...30H}
{Hesselbrock}, A.~J., \& {Minton}, D.~A. 2019, \bibinfo{title}{{Three Dynamical Evolution Regimes for Coupled Ring-satellite Systems and Implications for the Formation of the Uranian Satellite Miranda},} \aj, 157, 30, \dodoi{10.3847/1538-3881/aaf23a}

\bibitem[{K. {Hoch} {et~al.}(2025){Hoch}, {Rowland}, {Petrus}, {Nasedkin}, {Ingebretsen}, {Kammerer}, {Perrin}, {Rickman}, {D'Orazi}, {Kenworthy}, {Macintosh}, {Morley}, {De Rosa}, {Ruffio}, {Theissen}, {Konopacky}, {Ren}, {Chen}, {Gonzales}, {Chauvin}, {Bonnefoy}, {Palma-bifani}, {Girard}, {Pueyo}, {Moran}, {Ward-Duong}, {Balmer}, \& {Zhang}}]{Hoch2025}
{Hoch}, K., {Rowland}, M., {Petrus}, S., {et~al.} 2025, in American Astronomical Society Meeting Abstracts, Vol. 246, American Astronomical Society Meeting Abstracts, 413.03

\bibitem[{W.~B. {Hubbard} {et~al.}(1986){Hubbard}, {Brahic}, {Sicardy}, {Elicer}, {Roques}, \& {Vilas}}]{1986Natur.319..636H}
{Hubbard}, W.~B., {Brahic}, A., {Sicardy}, B., {et~al.} 1986, \bibinfo{title}{{Occultation detection of a neptunian ring-like arc},} \nat, 319, 636, \dodoi{10.1038/319636a0}

\bibitem[{R. {Hyodo} \& K. {Ohtsuki}(2015){Hyodo} \& {Ohtsuki}}]{2015NatGe...8..686H}
{Hyodo}, R., \& {Ohtsuki}, K. 2015, \bibinfo{title}{{Saturn's F ring and shepherd satellites a natural outcome of satellite system formation},} Nature Geoscience, 8, 686, \dodoi{10.1038/ngeo2508}

\bibitem[{M. {Kasper} {et~al.}(2021){Kasper}, {Cerpa Urra}, {Pathak}, {Bonse}, {Nousiainen}, {Engler}, {Heritier}, {Kammerer}, {Leveratto}, {Rajani}, {Bristow}, {Le Louarn}, {Madec}, {Str{\"o}bele}, {Verinaud}, {Glauser}, {Quanz}, {Helin}, {Keller}, {Snik}, {Boccaletti}, {Chauvin}, {Mouillet}, {Kulcs{\'a}r}, \& {Raynaud}}]{2021Msngr.182...38K}
{Kasper}, M., {Cerpa Urra}, N., {Pathak}, P., {et~al.} 2021, \bibinfo{title}{{PCS {\textemdash} A Roadmap for Exoearth Imaging with the ELT},} The Messenger, 182, 38, \dodoi{10.18727/0722-6691/5221}

\bibitem[{P.~C. {Keenan} \& R.~C. {McNeil}(1989){Keenan} \& {McNeil}}]{Keenan1989}
{Keenan}, P.~C., \& {McNeil}, R.~C. 1989, \bibinfo{title}{{The Perkins Catalog of Revised MK Types for the Cooler Stars},} \apjs, 71, 245, \dodoi{10.1086/191373}

\bibitem[{M.~A. {Kenworthy} \& E.~E. {Mamajek}(2015){Kenworthy} \& {Mamajek}}]{2015ApJ...800..126K}
{Kenworthy}, M.~A., \& {Mamajek}, E.~E. 2015, \bibinfo{title}{{Modeling Giant Extrasolar Ring Systems in Eclipse and the Case of J1407b: Sculpting by Exomoons?},} \apj, 800, 126, \dodoi{10.1088/0004-637X/800/2/126}

\bibitem[{P. {Kervella} {et~al.}(2003){Kervella}, {Th{\'e}venin}, {S{\'e}gransan}, {Berthomieu}, {Lopez}, {Morel}, \& {Provost}}]{Kervella2003}
{Kervella}, P., {Th{\'e}venin}, F., {S{\'e}gransan}, D., {et~al.} 2003, \bibinfo{title}{{The diameters of alpha Centauri A and B. A comparison of the asteroseismic and VINCI/VLTI views},} \aap, 404, 1087, \dodoi{10.1051/0004-6361:20030570}

\bibitem[{Y. {Liang} {et~al.}(2021){Liang}, {Robnik}, \& {Seljak}}]{2021AJ....161..202L}
{Liang}, Y., {Robnik}, J., \& {Seljak}, U. 2021, \bibinfo{title}{{Kepler-90: Giant Transit-timing Variations Reveal a Super-puff},} \aj, 161, 202, \dodoi{10.3847/1538-3881/abe6a7}

\bibitem[{T. {Lu} {et~al.}(2025){Lu}, {Li}, {Cassese}, \& {Lin}}]{2025ApJ...980...39L}
{Lu}, T., {Li}, G., {Cassese}, B., \& {Lin}, D.~N.~C. 2025, \bibinfo{title}{{The Dynamical History of HIP-41378 f{\textemdash}Oblique Exorings Masquerading as a Puffy Planet},} \apj, 980, 39, \dodoi{10.3847/1538-4357/ada4b2}

\bibitem[{M.~A. {Millar-Blanchaer} {et~al.}(2024){Millar-Blanchaer}, {Altinier}, {Carter}, {Choquet}, {De Rosa}, {Girard}, {Godoy}, {Hinkley}, {Leisenring}, {Macintosh}, {Pearce}, {Ray}, {Vancil}, \& {Vigan}}]{Millar-Blanchaer2024jwstprop}
{Millar-Blanchaer}, M.~A., {Altinier}, L., {Carter}, A., {et~al.} 2024, \bibinfo{title}{{Finding the great sculptors: A Renaissance in Planet Disk Dynamics},}, JWST Proposal. Cycle 3, ID. \#6012

\bibitem[{C.~D. {Murray} \& S.~F. {Dermott}(1999){Murray} \& {Dermott}}]{Murray1999}
{Murray}, C.~D., \& {Dermott}, S.~F. 1999, {Solar System Dynamics}, \dodoi{10.1017/CBO9781139174817}

\bibitem[{J. {O'Donoghue} {et~al.}(2016){O'Donoghue}, {Melin}, {Stallard}, {Provan}, {Moore}, {Badman}, {Cowley}, {Baines}, {Miller}, \& {Blake}}]{ODonoghue2016}
{O'Donoghue}, J., {Melin}, H., {Stallard}, T.~S., {et~al.} 2016, \bibinfo{title}{{Ground-based observations of Saturn's auroral ionosphere over three days: Trends in H$_{3}$$^{+}$ temperature, density and emission with Saturn local time and planetary period oscillation},} \icarus, 263, 44, \dodoi{10.1016/j.icarus.2015.04.018}

\bibitem[{M.~D. {Perrin} {et~al.}(2018{\natexlab{a}}){Perrin}, {Pueyo}, {Van Gorkom}, {Brooks}, {Rajan}, {Girard}, \& {Lajoie}}]{2018SPIE10698E..09P}
{Perrin}, M.~D., {Pueyo}, L., {Van Gorkom}, K., {et~al.} 2018{\natexlab{a}}, in Society of Photo-Optical Instrumentation Engineers (SPIE) Conference Series, Vol. 10698, Space Telescopes and Instrumentation 2018: Optical, Infrared, and Millimeter Wave, ed. M.~{Lystrup}, H.~A. {MacEwen}, G.~G. {Fazio}, N.~{Batalha}, N.~{Siegler}, \& E.~C. {Tong}, 1069809, \dodoi{10.1117/12.2313552}

\bibitem[{M.~D. {Perrin} {et~al.}(2018{\natexlab{b}}){Perrin}, {Pueyo}, {Van Gorkom}, {Brooks}, {Rajan}, {Girard}, \& {Lajoie}}]{Perrin2018}
{Perrin}, M.~D., {Pueyo}, L., {Van Gorkom}, K., {et~al.} 2018{\natexlab{b}}, in Society of Photo-Optical Instrumentation Engineers (SPIE) Conference Series, Vol. 10698, Space Telescopes and Instrumentation 2018: Optical, Infrared, and Millimeter Wave, ed. M.~{Lystrup}, H.~A. {MacEwen}, G.~G. {Fazio}, N.~{Batalha}, N.~{Siegler}, \& E.~C. {Tong}, 1069809, \dodoi{10.1117/12.2313552}

\bibitem[{A.~L. {Piro} \& S. {Vissapragada}(2020){Piro} \& {Vissapragada}}]{2020AJ....159..131P}
{Piro}, A.~L., \& {Vissapragada}, S. 2020, \bibinfo{title}{{Exploring Whether Super-puffs can be Explained as Ringed Exoplanets},} \aj, 159, 131, \dodoi{10.3847/1538-3881/ab7192}

\bibitem[{S.~P. {Quanz} {et~al.}(2013){Quanz}, {Amara}, {Meyer}, {Kenworthy}, {Kasper}, \& {Girard}}]{2013ApJ...766L...1Q}
{Quanz}, S.~P., {Amara}, A., {Meyer}, M.~R., {et~al.} 2013, \bibinfo{title}{{A Young Protoplanet Candidate Embedded in the Circumstellar Disk of HD 100546},} \apjl, 766, L1, \dodoi{10.1088/2041-8205/766/1/L1}

\bibitem[{B.~A. {Smith} {et~al.}(1979){Smith}, {Soderblom}, {Johnson}, {Ingersoll}, {Collins}, {Shoemaker}, {Hunt}, {Masursky}, {Carr}, {Davies}, {Cook}, {Boyce}, {Danielson}, {Owen}, {Sagan}, {Beebe}, {Veverka}, {Strom}, {McCauley}, {Morrison}, {Briggs}, \& {Suomi}}]{1979Sci...204..951S}
{Smith}, B.~A., {Soderblom}, L.~A., {Johnson}, T.~V., {et~al.} 1979, \bibinfo{title}{{The Jupiter System Through the Eyes of Voyager 1},} Science, 204, 951, \dodoi{10.1126/science.204.4396.951}

\bibitem[{B.~A. {Smith} {et~al.}(1989){Smith}, {Soderblom}, {Banfield}, {Barnet}, {Basilevksy}, {Beebe}, {Bollinger}, {Boyce}, {Brahic}, {Briggs}, {Brown}, {Chyba}, {Collins}, {Colvin}, {Cook}, {Crisp}, {Croft}, {Cruikshank}, {Cuzzi}, {Danielson}, {Davies}, {de Jong}, {Dones}, {Godfrey}, {Goguen}, {Grenier}, {Haemmerle}, {Hammel}, {Hansen}, {Helfenstein}, {Howell}, {Hunt}, {Ingersoll}, {Johnson}, {Kargel}, {Kirk}, {Kuehn}, {Limaye}, {Masursky}, {McEwen}, {Morrison}, {Owen}, {Owen}, {Pollack}, {Porco}, {Rages}, {Rogers}, {Rudy}, {Sagan}, {Schwartz}, {Shoemaker}, {Showalter}, {Sicardy}, {Simonelli}, {Spencer}, {Sromovsky}, {Stoker}, {Strom}, {Suomi}, {Synott}, {Terrile}, {Thomas}, {Thompson}, {Verbiscer}, \& {Veverka}}]{1989Sci...246.1422S}
{Smith}, B.~A., {Soderblom}, L.~A., {Banfield}, D., {et~al.} 1989, \bibinfo{title}{{Voyager 2 at Neptune: Imaging Science Results},} Science, 246, 1422, \dodoi{10.1126/science.246.4936.1422}

\bibitem[{J. {Szul{\'a}gyi} {et~al.}(2018){Szul{\'a}gyi}, {Cilibrasi}, \& {Mayer}}]{2018ApJ...868L..13S}
{Szul{\'a}gyi}, J., {Cilibrasi}, M., \& {Mayer}, L. 2018, \bibinfo{title}{{In Situ Formation of Icy Moons of Uranus and Neptune},} \apjl, 868, L13, \dodoi{10.3847/2041-8213/aaeed6}

\bibitem[{Y.~K. {Tang} \& N. {Gai}(2011){Tang} \& {Gai}}]{Tang2011}
{Tang}, Y.~K., \& {Gai}, N. 2011, \bibinfo{title}{{Asteroseismic modelling of the metal-poor star {\ensuremath{\tau}} Ceti},} \aap, 526, A35, \dodoi{10.1051/0004-6361/201014886}

\bibitem[{T.~I.~M. {van Werkhoven} {et~al.}(2014){van Werkhoven}, {Kenworthy}, \& {Mamajek}}]{vanWerkhoven2014}
{van Werkhoven}, T.~I.~M., {Kenworthy}, M.~A., \& {Mamajek}, E.~E. 2014, \bibinfo{title}{{Analysis of 1SWASP J140747.93-394542.6 eclipse fine-structure: hints of exomoons},} \mnras, 441, 2845, \dodoi{10.1093/mnras/stu725}

\bibitem[{A. {Vanderburg} {et~al.}(2016){Vanderburg}, {Becker}, {Kristiansen}, {Bieryla}, {Duev}, {Jensen-Clem}, {Morton}, {Latham}, {Adams}, {Baranec}, {Berlind}, {Calkins}, {Esquerdo}, {Kulkarni}, {Law}, {Riddle}, {Salama}, \& {Schmitt}}]{Vanderburg2016}
{Vanderburg}, A., {Becker}, J.~C., {Kristiansen}, M.~H., {et~al.} 2016, \bibinfo{title}{{Five Planets Transiting a Ninth Magnitude Star},} \apjl, 827, L10, \dodoi{10.3847/2041-8205/827/1/L10}

\bibitem[{J.~H. {Waite} {et~al.}(2018){Waite}, {Perryman}, {Perry}, {Miller}, {Bell}, {Cravens}, {Glein}, {Grimes}, {Hedman}, {Cuzzi}, {Brockwell}, {Teolis}, {Moore}, {Mitchell}, {Persoon}, {Kurth}, {Wahlund}, {Morooka}, {Hadid}, {Chocron}, {Walker}, {Nagy}, {Yelle}, {Ledvina}, {Johnson}, {Tseng}, {Tucker}, \& {Ip}}]{Waite2018}
{Waite}, J.~H., {Perryman}, R.~S., {Perry}, M.~E., {et~al.} 2018, \bibinfo{title}{{Chemical interactions between Saturn's atmosphere and its rings},} Science, 362, aat2382, \dodoi{10.1126/science.aat2382}

\bibitem[{X. {Wang} {et~al.}(2024){Wang}, {Li}, {Jiang}, {Fry}, {West}, {Nixon}, {Guan}, {Karandana G}, {Albright}, {Colwell}, {Guillot}, {Hofstadter}, {Kenyon}, {Mallama}, {Perez-Hoyos}, {Sanchez-Lavega}, {Simon}, {Wenkert}, \& {Zhang}}]{Wang2024}
{Wang}, X., {Li}, L., {Jiang}, X., {et~al.} 2024, \bibinfo{title}{{Cassini spacecraft reveals global energy imbalance of Saturn},} Nature Communications, 15, 5045, \dodoi{10.1038/s41467-024-48969-9}

\bibitem[{J. {Wisdom} {et~al.}(2022){Wisdom}, {Dbouk}, {Militzer}, {Hubbard}, {Nimmo}, {Downey}, \& {French}}]{2022Sci...377.1285W}
{Wisdom}, J., {Dbouk}, R., {Militzer}, B., {et~al.} 2022, \bibinfo{title}{{Loss of a satellite could explain Saturn{\textquoteright}s obliquity and young rings},} Science, 377, 1285, \dodoi{10.1126/science.abn1234}

\bibitem[{M. {Ygouf} {et~al.}(2024){Ygouf}, {Beichman}, {Llop-Sayson}, {Bryden}, {Leisenring}, {G{\'a}sp{\'a}r}, {Krist}, {Rieke}, {Rieke}, {Wolff}, {Roellig}, {Su}, {Hainline}, {Hodapp}, {Greene}, {Meyer}, {Kelly}, {Misselt}, {Stansberry}, {Boyer}, {Johnstone}, {Horner}, \& {Greenbaum}}]{Ygouf2024}
{Ygouf}, M., {Beichman}, C.~A., {Llop-Sayson}, J., {et~al.} 2024, \bibinfo{title}{{Searching for Planets Orbiting Fomalhaut with JWST/NIRCam},} \aj, 167, 26, \dodoi{10.3847/1538-3881/ad08c8}

\end{thebibliography}
\bibliographystyle{aasjournalv7}

\end{document}